\documentclass[iop,apj]{emulateapj}
\usepackage{epsf}

\newcommand{\etal}{{et al.~}}

\newcommand{\kms}{ \ {\rm{km \ s^{-1}}}\>}

\newcommand{\kpc}{\>{\rm kpc}}
\newcommand{\Myr}{\>{\rm Myr}}
\newcommand{\Msun}{\>{\rm M_{\odot}}}

\begin{document}

\pagenumbering{arabic}

\title{Third-Epoch Magellanic Cloud Proper Motions II:\\
The Large Magellanic Cloud Rotation Field in Three Dimensions}

\author{Roeland P.~van der Marel}
\affil{Space Telescope Science Institute, 3700 San Martin Drive, 
       Baltimore, MD 21218}
\author{Nitya Kallivayalil$^{1,2}$}
\affil{Yale Center for Astronomy \& Astrophysics,
       260 Whitney Ave, New Haven, CT}
\altaffiltext{1}{YCAA Prize Fellow}
\altaffiltext{2}{Dept. of Astronomy, University of
  Virginia, 530 McCormick Road, Charlottesville, VA 22904}

\slugcomment{ApJ, submitted, May 14, 2013}
\shorttitle{Large Magellanic Cloud Rotation Field in 3D}
\shortauthors{van der Marel \& Kallivayalil}

\begin{abstract}
We present the first detailed assessment of the large-scale rotation
of any galaxy based on full three-dimensional velocity
measurements. We do this for the Large Magellanic Cloud (LMC) by
combining our Hubble Space Telescope average proper motion (PM)
measurements for stars in 22 fields, with existing line-of-sight (LOS)
velocity measurements for 6790 individual stars. We interpret these
data with a model of circular rotation in a flat disk. The PM and LOS
data paint a consistent picture of the LMC rotation, and their
combination yields several new insights. The PM data imply a stellar
dynamical center that coincides with the HI dynamical center (but
offset from the photometric center), and a rotation curve amplitude
that is consistent with that inferred from LOS velocity studies. This
resolves several puzzles posed by existing work. The implied viewing
angles of the LMC disk agree with the range of values found in the
literature, but continue to indicate variations with stellar
population and/or radius in the disk. Young (red supergiant) stars
rotate faster than old (red and asymptotic giant branch) stars due to
asymmetric drift. Outside the central region, the rotation curve is
approximately flat out to the outermost data. The circular velocity
$V_{\rm circ} = 91.7 \pm 18.8 \kms$ (with the uncertainty dominated by
inclination uncertainties) is consistent with the baryonic
Tully-Fisher relation, and implies an enclosed mass $M(8.7 \kpc) =
(1.7 \pm 0.7) \times 10^{10} \Msun$. The virial mass is larger,
depending of the full extent of the LMC's dark halo. The tidal radius
is $22.3 \pm 5.2 \kpc$ ($24.0^{\circ} \pm 5.6^{\circ}$), if the
circular velocity stays flat this far out. Combination of the PM and
LOS data yields kinematic distance estimates for the LMC, but these
are not yet competitive with other methods.
\end{abstract}

\keywords{proper motions ---
galaxies: individual (Large Magellanic Cloud) ---
galaxies: kinematics and dynamics ---
Magellanic Clouds}

\section{Introduction}
\label{s:intro}

Measurements of galaxy rotation curves form the foundation of much of
our understanding of galaxy formation, structure, and dynamics (e.g.,
Binney \& Merrifield 1998; Binney \& Tremaine 2008; Mo, van den Bosch,
\& White 2010). The current knowledge of galaxy rotation is based
entirely on observations of Doppler shifts in radiation from
galaxies. This yields only one coordinate of motion, the LOS
velocity. If a galaxy rotates, and is not viewed edge-on, it will also
rotate in the plane of the sky. Until now, the implied PMs have
generally been undetectable, given the available observational
capabilities. However, the observational capabilities have steadily
advanced. We present here new results for the LMC that constitute the
first detailed measurement and analysis of the large-scale rotation
field of any galaxy in all three dimensions.\footnote{VLBI
  observations of water masers have been used to detect the PM
  rotation of {\it nuclear} gas disks in some galaxies (e.g., NGC
  4258; Herrnstein \etal 1999). Similar techniques can in principle be
  used to study the large-scale rotation curve of nearby galaxies
  (e.g., Brunthaler \etal 2005), but this has not yet been explored in
  detail.}

The Hubble Space Telescope (HST) provides a unique combination of high
spatial resolution, long-term stability, exquisite instrument
calibrations, and ever-increasing time baselines. Over the past
decade, this has opened up the Local Group of galaxies to detailed PM
studies. These studies have focused primarily on the satellites of the
Milky Way (Kallivayalil \etal 2006a, hereafter K06; Kallivayalil, van
der Marel \& Alcock 2006b; Piatek \& Prior 2008, and references
therein; Pryor, Piatek \& Olszewski 2010; L{\'e}pine \etal 2011; Sohn
\etal 2013; Boylan-Kolchin \etal 2013). More recently it has even
become possible to go out as far as M31 (Sohn \etal 2012; van der
Marel \etal 2012a,b). All of these studies have aimed at measuring the
systemic center-of-mass (COM) motion of the target galaxies, and not
their internal kinematics. So typically, only 1--3 different fields
were observed in any given galaxy. By contrast, a study of internal
kinematics requires, in addition to high PM accuracy, a larger number
of different fields spread out over the face of the galaxy.

In K06 we presented a detailed PM study of the LMC. We used HST to
observe 21 fields centered on background quasars, in two epochs
separated by a median baseline of 1.9 years. The distribution of
observed fields extends to 4$^{\circ}$ from the LMC center ($1^{\circ}
= 0.87 \kpc$ for an assumed distance of $50.1 \kpc$, i.e., $m-M =
18.50$; Freedman \etal 2001). From the data we derived the average PM
of the stars in each field. We used this to estimate the PM of the LMC
COM. In Besla \etal (2007) our team studied the implied orbit of the
Magellanic Clouds, and argued that they may be falling into the Milky
Way for the first time. The data also allowed us to detect the PM
rotation of the LMC at $1.3\sigma$ significance. The rotation sense
and magnitude were found to be consistent with the detailed
predictions for the LMC PM rotation field presented by van der Marel
\etal (2002; hereafter vdM02), based on the observed LOS rotation
field of carbon stars.

Piatek \etal (2008a, hereafter P08) performed a more sophisticated
reanalysis of our K06 data, including small corrections for
charge-transfer efficiency (CTE) losses. This yielded better PM
consistency between fields, but implied a similar PM for the LMC
COM. P08 used their measurements to derive the first crude PM rotation
curve for the LMC, assuming fixed values for the dynamical center and
disk orientation. However, their inferred rotation amplitude $V_{\rm
  rot} = 120 \pm 15 \kms$ appears unphysically high, exceeding the
known rotation of cold HI gas (Kim \etal 1998; Olsen \& Massey 2007)
by $\sim 40 \kms$. So better data are needed to accurately address the
PM rotation of the LMC.

We recently presented a third epoch of HST PM data for 10 fields
(Kallivayalil \etal 2013; hereafter Paper~I), increasing the median
time baseline to 7.1 years. For these fields we obtained a median
per-coordinate PM uncertainty of only 7 km/s (0.03 mas/yr), which is a
factor 3--4 better than in K06 and P08. This corresponds to $\sim
10$\% of the LMC rotation amplitude. As we show in the present paper,
these data are sufficient to map out the LMC PM rotation field in
detail, yielding new determinations of the LMC dynamical center, disk
orientation, and rotation curve.

The LMC is a particularly interesting galaxy for which to perform such
a study. At a distance of only $\sim 50 \kpc$, it is one of nearest
and best-studied galaxies next to our own Milky Way (e.g., Westerlund
1997; van den Bergh 2000). It is a benchmark for studies on various
topics, including stellar populations and the interstellar medium,
microlensing by dark objects, and the cosmological distance scale. As
nearby companion of the Milky Way, with significant signs of
interaction with the Small Magellanic Cloud (SMC), the LMC is also an
example of ongoing hierarchical structure formation. For all these
applications it is important to have a solid understanding of the LMC
structure and kinematics.

The current state of knowledge about the kinematics of the LMC was
reviewed recently by van der Marel, Kallivayalil \& Besla
(2009). Studies of the LOS velocities of many different tracers have
contributed to this knowledge.  The kinematics of gas in the LMC has
been studied primarily using HI (e.g., Kim \etal 1998; Olsen \& Massey
2007; Olsen \etal 2011, hereafter O11). Discrete LMC tracers which
have been studied kinematically include star clusters (e.g., Schommer
\etal 1992; Grocholski \etal 2006), planetary nebulae (Meatheringham
\etal 1988), HII regions (Feitzinger, Schmidt-Kaler \& Isserstedt
1977), red supergiants (Prevot \etal 1985; Massey \& Olsen 2003; O11),
red giant branch (RGB) stars (Zhao \etal 2003; Cole \etal 2005;
Carrera \etal 2011), carbon stars and other asymptotic giant branch
(AGB) stars (e.g., Kunkel \etal 1997; Hardy \etal 2001; vdM02; Olsen
\& Massey 2007; O11), and RR Lyrae stars (Minniti \etal 2003;
Borissova \etal 2006). For the majority of tracers, the line-of-sight
velocity dispersion is at least a factor $\sim 2$ smaller than their
rotation velocity. This implies that on the whole the LMC is a
(kinematically cold) disk system.

Specific questions that can be addressed in a new way through a study
of the LMC PM rotation field include the following:

\begin{itemize}

\item What is the stellar dynamical center of the LMC, and does this
  coincide with the HI dynamical center? It has long been known that
  different measures of the LMC center (e.g., center of the bar,
  center of the outer isophotes, HI dynamical center, etc.) are not
  spatially coincident (e.g., van der Marel 2001, hereafter vdM01;
  Cole \etal 2005), but a solid understanding of this remains lacking.

\item What is the orientation under which we view the LMC disk?  Past
  determinations of the inclination angle and the line-of-nodes
  position angle have spanned a significant range, and the results
  from different studies are often not consistent within the stated
  uncertainties (e.g., van der Marel \etal 2009). Knowledge of the
  orientation angles is necessary to determine the face-on properties
  of the LMC, with past work indicating that the LMC is not circular
  in its disk plane (vdM01).

\item What is the PM of the LMC COM, which is important for
  understanding the LMC orbit with respect to the Milky Way? We showed
  in Paper~I that the observational PM errors are now small enough
  that they are not the dominant uncertainty anymore. Instead,
  uncertainties in our knowledge of the geometry and kinematics of the
  LMC disk are now the main limiting factor.

\item What is the rotation curve amplitude of the LMC? Previous
  studies that used different tracers or methods sometimes obtained
  inconsistent values (e.g., P08; O11). The rotation
  curve amplitude is directly tied to the mass profile of the LMC,
  which is an important quantity for our understanding of the past
  orbital history of the LMC with respect to the Milky Way (Paper~I).

\item What is the distance of the LMC? Uncertainties in this distance
  form a key limitation in our understanding of the Hubble constant
  (e.g., Freedman \etal 2001). Comparison of the PM rotation amplitude
  (in mas/yr) and the LOS rotation amplitude (in km/s) can in
  principle yield a kinematical determination of the LMC distance that
  bypasses the stellar evolutionary uncertainties inherent to other
  methods (Gould 2000; van der Marel \etal 2009).

\end{itemize} 

In Paper~I of this series we presented our new third epoch
observations, and we analyzed all the available HST PM data for the
LMC (and the SMC). We included a reanalysis of the earlier K06/P08
data, with appropriate corrections for CTE losses. We used the data to
infer an improved value for the PM and the Galactocentric velocity of
the LMC COM, and we discussed the implications for the orbit of the
Magellanic Clouds with respect to the Milky Way (and in particular
whether or not the Clouds are on their first infall).

In the present paper we use the PM data from Paper~I to study the
internal kinematics of the LMC. The outline of this paper is as
follows. Section~\ref{s:model} discusses the PM rotation field,
including both the data and our best-fit model. Section~\ref{s:los}
presents a new analysis of the LOS kinematics of LMC tracers available
from the literature. By including the new constraints from the PM
data, this analysis yields a full three-dimensional view of the
rotation of the LMC disk. Section~\ref{s:param} discusses implications
of the results for our understanding of the geometry, kinematics, and
structure of the LMC. This includes discussions of the galaxy distance
and systemic motion, the dynamical center and rotation curve, the disk
orientation and limits on precession and nutation, and the galaxy
mass. We also discuss how the rotation of the LMC compares to that of
other galaxies. Section~\ref{s:conc} summarizes the main conclusions.

\section{Proper Motion Rotation Field}
\label{s:model}

\subsection{Data}
\label{ss:data}

We use the PM data presented in Table~1 of Paper~I as the basis of our
study. The data consist of positions $(\alpha,\delta)$ for 22 fields,
with measured PMs $(\mu_W,\mu_N)$ in the West and North directions,
and corresponding PM uncertainties $(\Delta \mu_W,\Delta
\mu_N)$. There are 10 ``high-accuracy'' fields with long time
baselines ($\sim 7$ years) and three-epochs of data\footnote{This
  includes one field with a long time baseline for which there is no
  data for the middle epoch.}, and 12 ``low-accuracy'' fields with
short time baselines ($\sim 2$ years) and two-epochs of data. The PM
measurement for each field represents the average PM of $N$ LMC stars
with respect to one known background quasar. The number of
well-measured LMC stars varies by field, but is in the range 8--129,
which a median $N=31$. The field size for each PM measurement
corresponds to the footprint of the HST ACS/HRC camera, which is $\sim
0.5 \times 0.5$ arcmin.\footnote{The third-epoch of data was obtained
  with the WFC3/UVIS camera, which has a larger field of
  view. However, the footprint of the final PM data is determined by
  the camera with the smallest field of view.} This is negligible
compared to the size of the LMC itself, which extends to a radius of
$10^{\circ}$--$20^{\circ}$ (vdM01; Saha \etal 2010).

\begin{figure*}[t]
\begin{center}
\epsfxsize=0.8\hsize
\centerline{\epsfbox{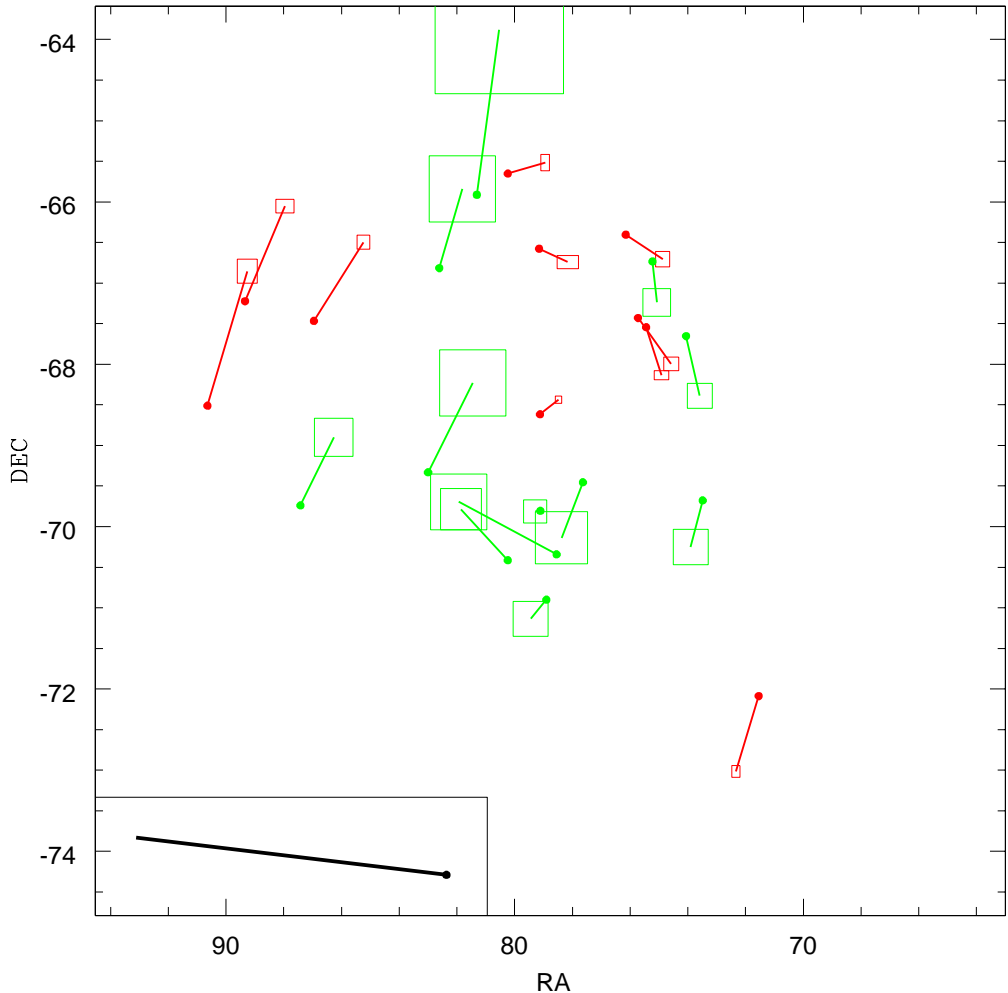}}
\caption{The spatially variable component ${\vec \mu}_{\rm obs,var}$
  of the observed LMC PM field. The positions of 22 fields observed
  with HST are indicated by solid dots. The PM vector shown for each
  field corresponds to the mean observed absolute PM of the stars in
  the given field, minus the constant vector ${\vec \mu}_0$ shown in
  the inset on the bottom left. The vector ${\vec \mu}_0$ is our
  best-fit for the PM of the LMC COM (see Table~\ref{t:param} and
  Paper~I). PMs are depicted by a vector that starts at the field
  location, with a size that (arbitrarily) indicates the mean
  predicted motion over the next $7.2 \Myr$. Clockwise motion is
  clearly evident. The uncertainty in each PM vector is illustrated by
  an open box centered on the end of each PM vector, which depicts the
  region $\pm \xi \Delta \mu_W$ by $\pm \xi \Delta \mu_N$. The
  constant $\xi = 1.36$ was chosen such that the box contains $68.3$\%
  of the two-dimensional Gaussian probability distribution.
  High-accuracy fields (with long time baselines, three epochs of
  data, and small error boxes) are shown in red, while low-accuracy
  fields (with short time baselines, two epochs of data, and larger
  error boxes) are shown in green. The figure shows an (RA,DEC)
  representation of the sky, with the horizontal and vertical extent
  representing an equal number of degrees on the sky. The figure is
  centered on the PM dynamical center $(\alpha_0,\delta_0)$ of the LMC,
  as derived in the present paper (see Table~\ref{t:param}).}
\label{f:obsvar}
\end{center}
\end{figure*}

Figure~\ref{f:obsvar} illustrates the data, by showing the spatially
variable component of the observed PM field, ${\vec \mu}_{\rm obs,
  var} \equiv {\vec \mu}_{\rm obs} - {\vec \mu}_0$, where the constant
vector ${\vec \mu}_0 = (\mu_{W0},\mu_{N0}) = (-1.9103, 0.2292)$
mas/yr. This vector is the best-fit PM of the LMC COM as derived later
in the present paper, and as discussed in Paper~I. Clockwise motion is
clearly evident. The goal of the subsequent analysis is to model this
motion to derive relevant kinematical and geometrical parameters for
the LMC.


\subsection{Velocity Field Model}
\label{ss:vmodel}

To interpret the LMC PM observations one needs a model for the PM
vector ${\vec \mu} = (\mu_W,\mu_N)$ as a function of position on the
sky. The PM model can be expressed as a function of equatorial
coordinates, ${\vec \mu}_{\rm mod}(\alpha,\delta)$, or as a function
of polar coordinates, ${\vec \mu}_{\rm mod}(\rho,\Phi)$, where $\rho$
is the angular distance from the LMC COM and $\Phi$ is the
corresponding position angle measured from North over East. Generally
speaking, the model can be written as a sum of two vectors, ${\vec
  \mu}_{\rm mod} = {\vec \mu}_{\rm sys} + {\vec \mu}_{\rm rot}$,
representing the contributions from the systemic motion of the LMC COM
and from the internal rotation of the LMC, respectively.

Consider first the contribution from the systemic motion. The
three-dimensional velocity that determines how the LMC COM moves
through space is a fixed vector. However, the projection of this
vector onto the West and North directions depends on where one looks
in the LMC. This introduces an important spatial variation in the PM
field, due to several different effects, including: (i) only a
fraction $\cos(\rho)$ of the LMC transverse velocity is seen in the PM
direction; (ii) a fraction $\sin(\rho)$ of the LMC LOS velocity is
also seen in the PM direction; and (iii) the directions of West and
North are not fixed in a zenithal projection centered on the LMC, due
to the deviation of $(\alpha,\delta)$ contours from an orthogonal grid
near the South Galactic pole (see figure~4 of van der Marel \& Cioni
2001, hereafter vdMC01). As a result, one can write ${\vec \mu}_{\rm
  sys}(\alpha,\delta) = {\vec \mu}_0 + {\vec \mu}_{\rm
  per}(\alpha,\delta)$. The first term is the constant PM of the LMC
COM, measured at the position of the COM. The second term is the
spatially varying component of the systemic contribution, which can be
referred to as the ``viewing perspective'' component.

To describe the component of internal rotation, we assume that the LMC
is a flat disk with circular streamlines. This is the same approach
that has been used successfully to model LOS velocities in the LMC
(e.g., vdM02; O11). However, it should be kept in mind
that this model is only approximately correct. The LMC is not circular
in its disk plane (vdM01), so the streamlines are not
expected to be exactly circular. Fortunately, the gravitational
potential is always rounder than the density distribution, so circular
streamlines should give a reasonable low-order approximation. Also,
the modest $V/\sigma$ of the LMC indicates that its disk is not
particularly thin (vdM02). So the flat-disk model should be viewed as
an approximation to the actual (three-dimensional) velocity field as
projected onto the disk plane.

At any point in the disk, the relation between the transverse velocity
$v_t$ in km/s and the PM $\mu$ in mas/yr is given by $\mu = v_t /
(4.7403885 D)$, where $D$ is the distance in kpc. The distance $D$ is
not the same for all fields, and is not the same as the distance $D_0$
of the LMC COM. The LMC is an inclined disk, so one side of the LMC is
closer to use than the other.  This has been quantified explicitly by
comparing the relative brightness of stars on opposite sides of the
LMC (e.g., vdMC01).

The analytical expressions for the PM field thus obtained, 
\begin{equation}
   {\vec \mu}_{\rm mod}(\alpha,\delta) =
      {\vec \mu}_{\rm 0} +
      {\vec \mu}_{\rm per}(\alpha,\delta)    +
      {\vec \mu}_{\rm rot}(\alpha,\delta)    ,
\label{pmfield}
\end{equation}
were presented in vdM02. We refer the reader to that paper for the 
details of the spherical trigonometry and linear algebra involved. 
The following model parameters uniquely define the model:

\begin{itemize}

\item The projected position $(\alpha_0,\delta_0)$ of the LMC COM, which is
  also the dynamical center of the LMC's rotation.

\item The orientation of the LMC disk, as defined by the inclination
  $i$ (with $0^{\circ}$ defined as face-on) and the position angle
  $\Theta$ of the line of nodes (the intersection of the disk and sky
  planes), measured from North over East. 

\item The PM of the LMC COM, $(\mu_{W0},\mu_{N0})$, 
  expressed in the heliocentric frame (i.e., not corrected for the
  reflex motion of the Sun).

\item The heliocentric LOS velocity of the LMC COM, $v_{\rm
  LOS,0}/D_0$, expressed in angular units (for which we use mas/yr
  throughout this paper).

\item The rotation curve in the disk, $V(R')/D_0$, expressed in
  angular units. Here $R$ is the radius in the disk in physical units,
  and $R' \equiv R/D_0$. (Along the line of nodes, $R' = \tan(\rho)$;
  in general, the LMC distance must be specified to calculate the
  radius in the disk is in physical units).

\end{itemize}

\noindent The first two bullets define the geometrical properties of
the LMC, and the last three bullets its kinematical properties.

Figures~10a,b of vdM02 illustrate the predicted morphology of the PM
fields $\mu_{\rm per}$ and $\mu_{\rm rot}$ for a specific LMC model
tailored to fit the LOS velocity field. These two components have
comparable amplitudes. The spatially variable component of the
observed PM field ${\vec \mu}_{\rm obs, var}$ in Figure~\ref{f:obsvar}
provides an observational estimate of the sum ${\vec \mu}_{\rm per} +
{\vec \mu}_{\rm rot}$ (compare eq.~[\ref{pmfield}]).

\subsection{Information Content of the Proper Motion and Line-of-Sight 
Velocity Fields}
\label{ss:pmvslos}

The PM field is defined by the variation of two components of motion
over the face of the LMC. By contrast, the LOS velocity field is
defined by the variation of only one component of motion. The PM field
therefore contains more information, and has more power to
discriminate the parameters of the model. As we will show, important
constraints can be obtained with only 22 PM
measurements,\footnote{Bekki (2011) argued incorrectly that PM
  observations for up to 1000 quasar fields would be required to
  obtain meaningful constraints. In his simulations, he assumed that
  only one particle is measured per field. In practice, we measure
  many stars per field (median $N=31$), so the shot noise is a factor
  $\sim \sqrt{31}$ smaller. Moreover, his simulation had a central
  velocity dispersion in the disk of $\sigma = 50 \kms$, which is
  higher than the value $\sigma = 20 \kms$ typical for the population
  that dominates the mass (vdM02). As a result, his assumed random
  measurement uncertainties were a factor $\sim 14$ too high.}
whereas LOS velocity studies require hundreds or thousands of stars.

The following simple arguments show that knowledge of the full PM
field in principle allows unique determination of all model
parameters, without degeneracy.

\begin{itemize} 

\item The dynamical center $(\alpha_0,\delta_0)$ is the position
  around which the spatially variable component of the PM field has a
  well-defined sense of rotation.

\item The azimuthal variation of the PM rotation field determines both
  of the LMC disk orientation angles $(\Theta,i)$. Perpendicular to
  the line of nodes (i.e., $\Phi = \Theta \pm 90^{\circ}$), all of the
  rotational velocity $V(R')$ in the disk is seen as a PM (and none is
  seen along the LOS). By contrast, along the line of nodes (i.e.,
  $\Phi = \Theta$ or $\Theta + 180^{\circ}$), only approximately $V(R')
  \cos i$ is seen as a PM (and approximately $V(R') \sin i$ is seen
  along the LOS). The near and far side of the disk are distinguished
  by the fact that velocities on the near side imply larger PMs.

\item The PM of the LMC COM, $(\mu_{W0},\mu_{N0})$, is the PM at the
  dynamical center.

\item The systemic LOS velocity $v_{\rm LOS,0}/D_0$ in angular units
  follows from the radially directed component of the PM field. A
  fraction $\sin(\rho) v_{\rm LOS,0}$ is seen in this direction
  (appearing as an ``inflow'' for $v_{\rm LOS,0} > 0$ and an
  ``outflow'' for $v_{\rm LOS,0} < 0$). This component is almost
  perpendicular to the more tangentially oriented component induced by
  rotation in the LMC disk, so the two are not degenerate. However,
  the radially directed component is small near the galaxy center
  (e.g., $\sin(\rho) \lesssim 0.07$ for $\rho \lesssim 4^{\circ}$), so
  exquisite PM data would be required to constrain $v_{\rm LOS,0}/D_0$
  with meaningful accuracy.

\item The rotation curve $V(R')/D_0$ in angular units follows from the
  PMs along the line-of-nodes position angle $\Theta$.

\end{itemize}


\begin{deluxetable*}{llrrr}
\setlength{\tabcolsep}{20pt}
\setlength{\tablewidth}{\hsize}
\tablecaption{LMC Model Parameters: New Fit Results from Three-Dimensional 
Kinematics\label{t:param}}
\tablehead{
\colhead{Quantity} & \colhead{Unit} & \colhead{PMs} & \colhead{PMs+Old} & \colhead{PMs+Young}\\
\colhead{} & \colhead{} & \colhead{} & \colhead{$v_{\rm LOS}$ sample}
 & \colhead{$v_{\rm LOS}$ sample}\\
\colhead{(1)} & \colhead{(2)} & \colhead{(3)} & \colhead{(4)} & \colhead{(5)}}
\startdata
$\alpha_0$          &  deg    & $78.76 \pm 0.52$ 
                              & $79.88 \pm 0.83$                    
                              & $80.05 \pm 0.34$\\
$\delta_0$          &  deg    & $-69.19 \pm  0.25$
                              & $-69.59 \pm 0.25$
                              & $-69.30 \pm 0.12$\\
$i$                 &  deg    & $ 39.6 \pm 4.5$     
                              & $ 34.0 \pm 7.0$
                              & $ 26.2 \pm 5.9$\\
$\Theta$            &  deg    & $147.4 \pm 10.0$                   
                              & $139.1 \pm  4.1$
                              & $154.5 \pm  2.1$\\
$\mu_{W0}$          &  mas/yr & $-1.910 \pm 0.020$   
                              & $-1.895 \pm 0.024$
                              & $-1.891 \pm 0.018$\\
$\mu_{N0}$          &  mas/yr & $0.229 \pm 0.047$   
                              & $0.287 \pm 0.054$
                              & $0.328 \pm 0.025$\\
$v_{\rm LOS,0}$     &  km/s   & $262.2 \pm 3.4$\tablenotemark{[1]}
                              & $261.1 \pm 2.2$
                              & $269.6 \pm 1.9$\\
$V_{0,{\rm PM}}/D_0$  &  mas/yr & $0.320 \pm 0.029$ 
                              & $0.353 \pm 0.034$
                              & $0.289 \pm 0.025$\\
$V_{0,{\rm PM}}$\tablenotemark{[2]}
                    &  km/s   & $76.1 \pm  7.6$
                              & $83.8 \pm  9.0$
                              & $68.8 \pm  6.4$\\
$V_{0,{\rm LOS}}$   &  km/s   & $\ldots$ 
                              & $55.2 \pm 10.3$
                              & $89.3 \pm 18.8$\\
$V_{0,{\rm LOS}}\>\sin i$\tablenotemark{[2]} 
                    &  km/s  & $\ldots$ 
                              & $30.9 \pm 2.6$
                              & $39.4 \pm 1.9$\\
$R_0/D_0$           &         & $0.024 \pm 0.010$
                              & $0.075 \pm 0.005$
                              & $0.040 \pm 0.003$\\
$D_0$\tablenotemark{[3]}
                    & kpc     & $50.1 \pm 2.5 \kpc$
                              & $50.1 \pm 2.5 \kpc$
                              & $50.1 \pm 2.5 \kpc$\\
\enddata

\tablecomments{Column~(1) lists the model quantity, and column~(2) its
  units. Column~(3) lists the values from the model fit to the PM data
  in Section~\ref{s:model}. Columns~(4) and~(5) list the values from
  the model fit to the combined PM and LOS velocity data in
  Section~\ref{s:los}, for the old and young $v_{\rm LOS}$ sample,
  respectively. From top to bottom, the following quantities are
  listed: Position $(\alpha_0,\delta_0)$ of the dynamical center;
  Orientation angles $(i,\Theta)$ of the disk plane, being the
  inclination angle and line-of-nodes position angle, respectively; PM
  $(\mu_{W0},\mu_{N0})$ of the COM; LOS velocity $v_{\rm LOS,0}$ of
  the COM; Amplitude $V_{0,{\rm PM}}/D_0$ or $V_{0,{\rm PM}}$ of the
  rotation curve in angular units or physical units, respectively, as
  inferred from the PM data. Amplitude $V_{0,{\rm LOS}}$ of the
  rotation curve as inferred from the LOS velocity data, and observed
  component $V_{0,{\rm LOS}}\>\sin i$. Turnover radius $R_0/D_0$ of
  the rotation curve, expressed as a fraction of the distance (the
  rotation curve being parameterized so that it rises linearly to
  velocity V0 at radius R0, and then stays flat at larger radii); and
  the distance $D_0$.}

\tablenotetext{[1]}{Value from vdM02, not
  independently determined by the model fit. Uncertainty propagated into 
  all other model parameters.}
\tablenotetext{[2]}{Quantity derived from other parameters, accounting for 
  correlations between uncertainties.}
\tablenotetext{[3]}{Value from Freedman \etal (2001), corresponding to a 
  distance modulus $m-M = 18.50 \pm 0.10$, not
  independently determined by the model fit. Uncertainty propagated into
  all other model parameters.}
\end{deluxetable*}


\begin{figure*}[t]
\begin{center}
\epsfxsize=0.8\hsize
\centerline{\epsfbox{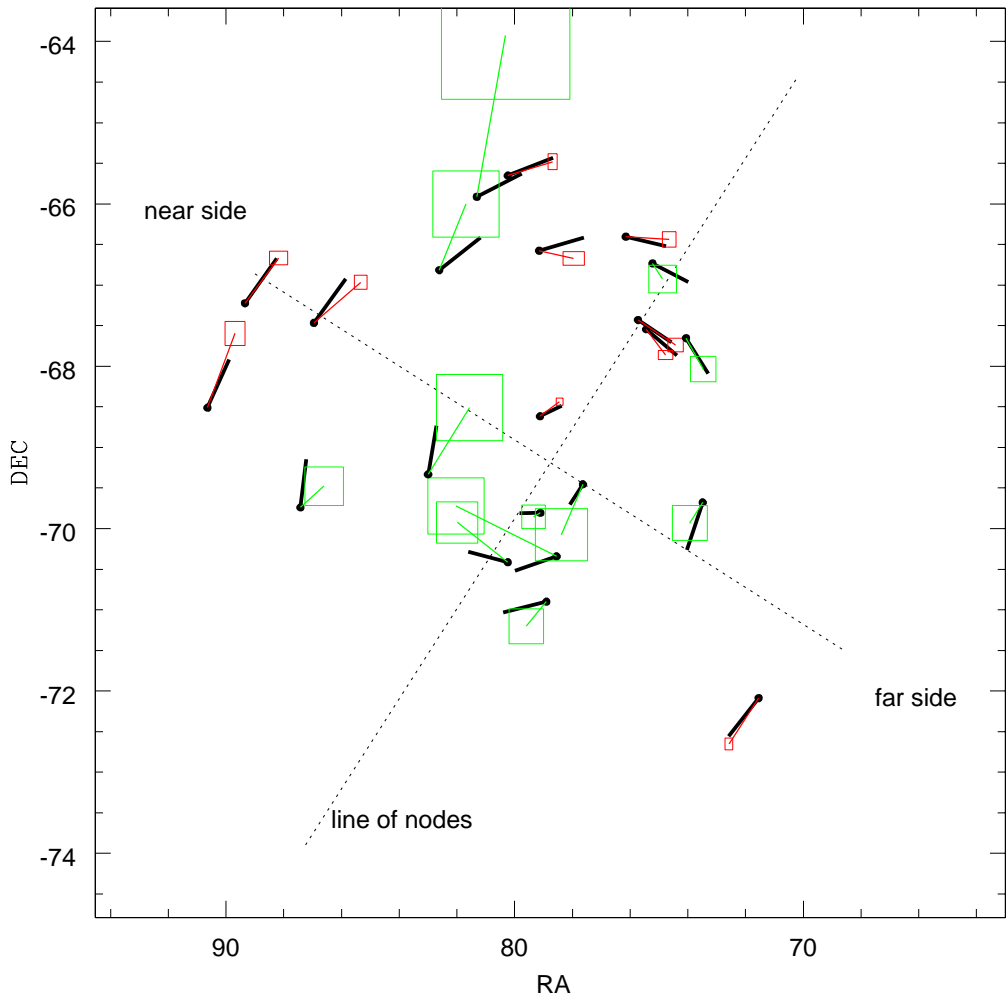}}
\caption{Data-model comparison for the rotation component ${\vec
    \mu}_{\rm obs,rot}$ of the observed LMC PM field, with similar
  plotting conventions as in Figure~\ref{f:obsvar}. For each field we
  now show in color the mean observed absolute PM of the stars in the
  given field, minus the component ${\vec \mu}_{\rm sys} = {\vec
    \mu}_0 + {\vec \mu}_{\rm per}$ implied by the best-fit model (see
  Table~\ref{t:param}). The latter subtracts the systemic motion of
  the LMC, and includes not only the PM of the LMC COM (as in
  Figure~\ref{f:obsvar}) but also the spatially-varying viewing
  perspective component. Solid black vectors show the rotation
  component ${\vec \mu}_{\rm rot}$ of the best-fit model. The
  observations show clockwise motion, which is reproduced by the
  model. A dotted line indicates the line of nodes, along position
  angle $\Theta$. Another dotted line connects the near and the far
  sides of the LMC disk, along position angles $\Theta-90^{\circ}$ and
  $\Theta+90^{\circ}$, respectively. Along the near-far direction, PMs
  are larger by a factor $1 / \cos i$ than along the line of
  nodes. However, distances along the near-far direction are
  foreshortened by a factor $\cos i$ compared to distances along the
  line of nodes (as indicated by the length of the dotted lines). The lines
  intersect at the dynamical center $(\alpha_0,\delta_0)$. The
  geometrical parameters $(\Theta, i, \alpha_0,\delta_0)$ are all
  uniquely defined by the model fit to the data, as is the rotation
  curve in the disk which is shown in Figure~\ref{f:rotcurve}.}
\label{f:obsrot}
\end{center}
\end{figure*}

By contrast, full knowledge of the LOS velocity field does {\it not}
constrain all the model parameters uniquely. Specifically, there is
strong degeneracy between three of the model parameters (see vdM02):
the rotation curve $V(R')$, the inclination angle $i$ (since the
observed LOS velocity component is approximately $V(R') \sin i$), and
the component $v_{t0c}$ of the transverse COM velocity vector ${\vec
  v}_{t0}$ projected onto the line of nodes (which adds a solid-body
component to the observed rotation). So the rotation curve can only be
determined from the LOS velocity field if $i$ and $v_{t0c}$ are
assumed to be known independently. Typically (e.g., vdM02; O11), $i$
has been estimated from geometric methods (e.g., vdMC01) and $v_{t0c}$
from proper motion studies (e.g., K06). It should be noted that the
transverse COM velocity component $v_{t0s}$ in the direction
perpendicular to the line of nodes {\it is} determined uniquely by the
LOS velocity field, as is the position angle $\Theta$ of the line of
nodes itself. And of course, the systemic LOS velocity $v_{\rm LOS,0}$
is determined much more accurately by the LOS velocity field than by
the PM field.

An important difference between the two observationally accessible
fields is that the PM field constrains velocities in angular units
(mas/yr), whereas the LOS velocity field constrains the same
velocities in physical units (km/s). Hence, comparison of the results
for e.g. $V(R')$ or $v_{t0s}$ from the two fields constrains the LMC
distance $D_0$. This is discussed further in Section~\ref{ss:dist}.
 
\subsection{Fitting Methodology}
\label{ss:method}

In our earlier analysis of K06, we treated $(\mu_{W0},\mu_{N0})$ as
the only free parameters to be determined from the PM data. All other
quantities were kept fixed to estimates previously obtained either by
vdM02 from a study of the LMC LOS velocity field, by vdMC01 from a
study of the LMC orientation angles, or by Freedman \etal (2001) from
a study of the LMC distance. P08 took the same approach, but as
discussed in Section~\ref{s:intro}, they did treat the rotation curve
$V(R')$ as a free function to be determined from the data. Keeping
model parameters fixed a priori is reasonable when only limited data
is available. However, this does have several undesirable
consequences. First, it does not use the full information content of
the PM data, which actually constrains the parameters
independently. Second, it opens the possibility that parameters are
used that are not actually consistent with the PM data. And third, it
leads to underestimates of the errorbars on the LMC COM PM
$(\mu_{W0},\mu_{N0})$, since the uncertainties in the geometry and
rotation of the LMC are not propagated into the answers (as discussed
in Paper~I).

The three-epoch PM data presented in Paper~I have much improved
quality over the two-epoch measurements presented by K06a and P08, as
evident from Figure~\ref{f:obsvar}. We therefore now treat all of the
key parameters that determine the geometry and kinematics of the LMC
as free parameters to be determined from the data. There are $M=22$ LMC
fields, and hence $N_{\rm data} = 2M = 44$ observed quantities (there are
two PM coordinates per field). By comparison, the model is defined by
the 7 parameters $(\alpha_0,\delta_0,\mu_{W0},\mu_{N0},v_{\rm
  LOS,0}/D_0,i,\Theta)$ and the one-dimensional function $V(R')/D_0$.
The rotation curves of galaxies follow well-defined patterns, and are
therefore easily parameterized with a small number of parameters. We use
a very simple form with two parameters
\begin{equation}
  V(R')/D_0 = (V_0/D_0) \> \min[R'/(R_0/D_0), \> 1)] 
\label{rotparam}
\end{equation}
(similar to P08 and O11). This corresponds to a rotation curve that
rises linearly to velocity $V_0$ at radius $R_0$, and stays flat
beyond that. The quantity $V_0/D_0$ is the rotation amplitude
expressed in angular units. Later in Section~\ref{ss:rotcurve} we also
present unparameterized estimates of the rotation curve
$V(R')$. Sticking with the parameterized form for now, we have an
overdetermined problem with more data points ($N_{\rm data} = 44$)
than model parameters ($N_{\rm param} = 9$), so this is a well-posed
mathematical problem. We also know from the discussion in
Section~\ref{ss:pmvslos} that the model parameters should be uniquely
defined by the data without degeneracy. So we proceed by numerical
fitting of the model to the data.

To fit the model we define a $\chi^2$ quantity 
\begin{eqnarray}
  \chi^2_{\rm PM} \equiv
  \sum_{i=1}^{M} & & \left [ (\mu_{W,{\rm obs},i} - \mu_{W,{\rm mod},i}) / 
                         \Delta \mu_{W,{\rm obs},i} \right ]^2 + \nonumber \\
                 & & \left [ (\mu_{N,{\rm obs},i} - \mu_{N,{\rm mod},i}) / 
                         \Delta \mu_{N,{\rm obs},i} \right ]^2       
\label{chisqpm}
\end{eqnarray}
that sums the squared residuals over all $M$ fields. We minimize
$\chi^2_{\rm PM}$ as function of the model parameters using a
down-hill simplex routine (Press \etal 1992). Multiple iterations and
checks were built in to ensure that a global minimum was found in the
multi-dimensional parameter space, instead of a local minimum.


\begin{deluxetable*}{llrr}
\setlength{\tabcolsep}{20pt}
\setlength{\tablewidth}{\hsize}
\tablecaption{LMC Model Parameters: Literature Results from Line-of-Sight 
Velocity Analyses\label{t:paramlit}}
\tablehead{
\colhead{Quantity} & \colhead{Unit} & \colhead{vdM02} & \colhead{O11}\\
 & & \colhead{(carbon stars)} & \colhead{(RSGs)} \\
\colhead{e(1)} & \colhead{(2)} & \colhead{(3)} & \colhead{(4)} }
\startdata
$\alpha_0$          &  deg    & $81.91 \pm 0.98$                    
                              & $81.91 \pm 0.98$\tablenotemark{[1,2,b]}\\
$\delta_0$          &  deg    & $-69.87 \pm 0.41$
                              & $-69.87 \pm 0.41$\tablenotemark{[1,2,b]}\\
$i$                 &  deg    & $ 34.7 \pm 6.2$\tablenotemark{[1,a]} 
                              & $ 34.7 \pm 6.2$\tablenotemark{[1,2,a]}\\
$\Theta$            &  deg    & $129.9 \pm  6.0$
                              & $142   \pm 5$\\
$\mu_{W0}$          &  mas/yr & $-1.68  \pm 0.16$\tablenotemark{[1,c]} 
                              & $-1.956 \pm 0.036$\tablenotemark{[1,2,d]} \\
$\mu_{N0}$          &  mas/yr & $0.34  \pm 0.16$\tablenotemark{[1,c]}
                              & $0.435 \pm 0.036$\tablenotemark{[1,2,d]} \\
$v_{\rm LOS,0}$     &  km/s   & $262.2 \pm 3.4$
                              & $263   \pm 2$ \\
$V_{0,{\rm LOS}}$   &  km/s   & $49.8 \pm 15.9$
                              & $87 \pm 5$\tablenotemark{[5,6]} \\
$V_{0,{\rm LOS}}\>\sin i$\tablenotemark{[3]} 
                    &  km/s   & $28.4 \pm 7.9$
                              & $50 \pm 3$\tablenotemark{[6]}\\
$R_0/D_0$           &         & $0.080 \pm 0.004$\tablenotemark{[4]}
                              & $0.048 \pm 0.002$ \\
$D_0$               & kpc     & $50.1 \pm 2.5 \kpc$\tablenotemark{[1,7]}
                              & $50.1 \pm 2.5 \kpc$\tablenotemark{[1,2,7]} \\
\enddata

\tablecomments{Parameters from model fits to LMC LOS velocity data, as
  obtained by vdM02 and O11; listed in columns (3) and (4),
  respectively. The table layout and the quantities in column~(1) are
  as in Table~\ref{t:param}. Parameter uncertainties are from the
  listed papers. Many of these are underestimates, for the reasons
  stated in the footnotes.}

\tablenotetext{[1]}{Value from a different source, not
  independently determined by the model fit.}
\tablenotetext{[2]}{Uncertainties in this parameter were not
  propagated in the model fit.}  
\tablenotetext{[3]}{Quantity derived from other parameters.}
\tablenotetext{[4]}{Determined by fitting a function
  of the form in equation~(\ref{rotparam}) to Table~2 of vdM02.}
\tablenotetext{[5]}{Degenerate with $\sin i$. The uncertainty is an
  underestimate. It does not reflect the listed inclination uncertainty,
  which adds an uncertainty of 15.6\% to $V_{0,{\rm LOS}}$.}
\tablenotetext{[6]}{Degenerate with $\mu_{c0} \equiv - \mu_{W0}
  \sin \Theta + \mu_{N0} \cos \Theta$. The uncertainty is an
  underestimate, and does not reflect the listed uncertainty in the COM PM, 
  or the use of now outdated values for the COM PM.}
\tablenotetext{[7]}{Value from Freedman \etal (2001), corresponding to a 
  distance modulus $m-M = 18.50 \pm 0.10$, not
  independently determined by the model fit.}


\tablenotetext{[a]}{vdM0C1.}
\tablenotetext{[b]}{vdM02.}
\tablenotetext{[c]}{average of pre-HST measurements compiled in vdM02.}
\tablenotetext{[d]}{P08.}
\end{deluxetable*}

Once the best-fitting model parameters are identified, we calculate
error bars on the model parameters using Monte Carlo simulations. Many
different pseudo–data sets are created that are analyzed similarly to
the real data set. The dispersions in the inferred model parameters
are a measure of the $1\sigma$ random errors on the model parameters.
Each pseudo–data set is created by calculating for each observed field
the best-fit model PM prediction, and by adding to this random
Gaussian deviates. The deviates are drawn from the known observational
error bars, multiplied by a factor $(\chi^2_{\rm
  min}/N_{DF})^{1/2}$. Here $\chi^2_{\rm min}$ is the $\chi^2_{\rm
  PM}$ value of the best-fit model, and $N_{\rm DF} = N_{\rm data} -
N_{\rm param} + N_{\rm fixed}$ is the number of degrees of freedom,
with $N_{\rm fixed}$ the number of parameters (if any) that are not
optimized in the fit. In practice we find that $\chi^2_{\rm min}$ is
somewhat larger than $N_{DF}$, indicating that the actual scatter in
the data is slightly larger than what is accounted for by random
errors. This is not surprising, given the complexity of the
astrometric data analysis and the relative simplicity of the
model. The approach used to create the pseudo-data ensures that the
actual scatter is propagated into the final uncertainties on the model
parameters.

It is known from LOS velocity studies that $v_{\rm LOS,0} = 262.2 \pm
3.4 \kms$ (vdM02), and from stellar population studies that $D_0 =
50.1 \pm 2.5$ kpc ($m-M = 18.50 \pm 0.10$; Freedman \etal
2001\footnote{The more recent study of Freedman \etal (2012) obtained
  a smaller uncertainty, $m-M = 18.477 \pm 0.033$, but to be
  conservative, we use the older Freedman \etal (2001) distance
  estimate throughout this paper.}). So $v_{\rm LOS,0}$ is known to
$\sim 1\%$ accuracy and $D$ to $\sim 5\%$ accuracy. Not surprisingly,
we have found that the PM data cannot constrain the model parameter
$v_{\rm LOS,0}/D_0$ with similar accuracy. Therefore, we have kept
$v_{\rm LOS,0}/D_0$ fixed in our analysis to the value implied by
existing knowledge. At $m-M = 18.50$, 1 mas/yr corresponds to $237.58
\kms$. Hence, $v_{\rm LOS,0}/D_0 = 1.104 \pm 0.053$ mas/yr. The
uncertainty in this value was propagated into the analysis by using
randomly drawn $v_{\rm LOS,0}/D_0$ values in the fitting of the
different Monte-Carlo generated pseudo-data sets.

\subsection{Data-Model Comparison}
\label{ss:compare}

Table~\ref{t:param} lists the parameters of the best-fit model and
their uncertainties. These parameters are discussed in detail in
Section~\ref{s:param}. Figure~\ref{f:obsrot} shows the data-model
comparison for the best fit. For this figure, we subtracted the
systemic velocity contribution ${\vec \mu}_{\rm sys} = {\vec \mu}_0 +
{\vec \mu}_{\rm per}$ implied by the best-fit model, from both the
observations and the model. By contrast to Figure~\ref{f:obsvar}, this
now also subtracts the spatially-varying viewing perspective. So the
observed rotation component ${\vec \mu}_{\rm obs,rot} \equiv {\vec
  \mu}_{\rm obs} - {\vec \mu}_0 - {\vec \mu}_{\rm per}$ is compared to
the model component ${\vec \mu}_{\rm rot}$. Clockwise motion is
clearly evident in the observations, and this is reproduced by the
model.

The best-fit model has $\chi^2_{\rm min} = 116.0$ for $N_{DF} =
36$. Hence, $(\chi^2_{\rm min}/N_{DF})^{1/2} = 1.80$. So even though
the model captures the essence of the observations, it is not formally
statistically consistent with it. There are three possible
explanations for this. First, the observations could be affected by
unidentified low-level systematics in the data analysis, in addition
to the well-quantified random uncertainties. Second, shot noise from
the finite number of stars may be important for some fields with low
$N$, causing the mean PM of the observed stars to deviate from the
true mean motion in the LMC disk. And third, the model may be too
over-simplified (e.g., if there are warps in the disk, or if the
streamlines in the LMC disk deviate from circles at a level comparable
to our measurement uncertainties). It is difficult to establish which
explanation may be correct, and the explanation may be different for
different fields.

Two of our HST fields are close to each other at a separation of only
$0.16^{\circ}$, and this provides some additional insight into
potential sources of error. The fields, labeled L12 and L14 in table~1
of Paper~I, are located at $\alpha \approx 75.6^{\circ}$ and $\delta
\approx -67.5^{\circ}$ (see Figure~\ref{f:obsvar}). Since the fields
are so close to each other, the best-fit model predicts that the PMs
should be similar, ${\vec \mu}_{\rm mod,L12} - {\vec \mu}_{\rm
  mod,L14} = (-0.015, -0.031)$ mas/yr. However, the observations
differ by ${\vec \mu}_{L12} - {\vec \mu}_{L14} = (-0.110 \pm 0.047 ,
-0.001 \pm 0.037)$ mas/yr. This level of disagreement can in principle
happen by chance (9\% probability), but maybe a possible additional
source of error is to blame. The disagreement in this case cannot
arise because the model is too oversimplified, since almost any model
would predict that closely-separated fields in the disk have similar
PMs. Also, shot noise is too small to explain the difference.  These
fields had $N=16$--18 stars measured, and a typical velocity
dispersion in the disk is $\sigma \approx 20 \kms$ (vdM02). This
implies a shot noise error (per coordinate, per field) of only $\sim
0.02$ mas/yr, which is below the random errors for these fields.
These fields have lower $N$ and smaller random errors than most other
fields, so this means that shot noise in general plays at most a small
role. So in the case of these fields, and maybe for the sample as a
whole, it is likely that we are dealing with unidentified low-level
systematics in the data analysis.

Either way, the exact cause why $\chi^2_{\rm min} > N_{DF}$ does not
matter much, since in the Monte-Carlo analysis of pseudo-data we
multiply all observational errors by $(\chi^2_{\rm
  min}/N_{DF})^{1/2}$. So the actual residuals in the data-model
comparison are accounted for when calculating the uncertainties in the
model parameters. Moreover, the astrometric observations presented in
Paper~I are extremely challenging. So the level of agreement in the
data-model comparison of Figure~\ref{f:obsrot} is actually extremely
encouraging.

\section{Line-of-Sight Rotation Field}
\label{s:los}

Many studies exist of the LOS velocity field of tracers in the LMC, as
discussed in Section~\ref{s:intro}. Two of the most sophisticated
studies are those of vdM02 and O11. The vdM02 study modeled the LOS
velocities of $\sim 1000$ carbon stars, and its results formed the
basis of the rotation model used in K06. The more recent O11 study
obtained a rotation fit to the LOS velocities of $\sim 700$ red
supergiants (RSGs), and also presented $\sim 4000$ new LOS velocities
for other giant and AGB stars. The parameters of the vdM02 and O11
rotation models are presented in Table~\ref{t:paramlit}.

Comparison of the vdM02 and O11 parameters to those obtained from our
PM field fit in Table~\ref{t:param} shows a few important
differences. The COM PM values used by both vdM02 and O11 are
inconsistent with our most recent estimate from Paper~I. This is
important, because the transverse motion of the LMC introduces a solid
body rotation component into the LMC LOS velocity field, which must be
corrected to model the internal LMC rotation. Also, the dynamical
centers either inferred (vdM02) or used (O11) by the past LOS velocity
studies are in conflict with the dynamical center implied by the new
PM analysis. These differences are discussed in detail in
Section~\ref{s:param}. Motivated by these differences, we decided to
perform a new analysis of the available LOS velocity data from the
literature, taking into account the new PM results. This yields a full
three-dimensional view of the rotation of the LMC disk.


\begin{figure*}
\begin{center}
\epsfxsize=0.8\hsize
\centerline{\epsfbox{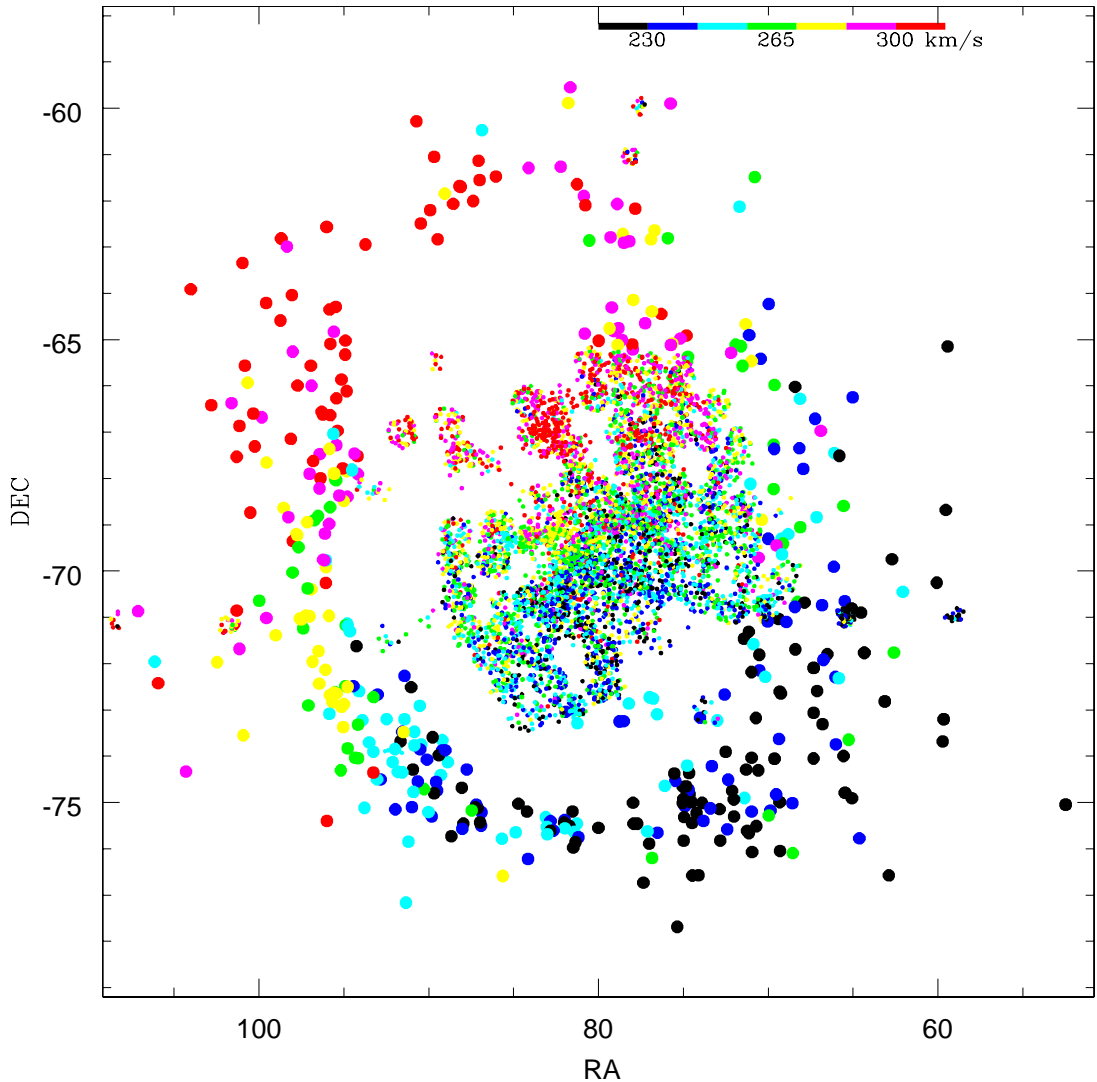}}
\caption{LMC LOS velocity field defined by 6790 observed stellar
  velocities available from the literature. All stars in the combined
  young and old samples discussed in the text are shown. Each star is
  color-coded by its velocity according to the legend at the top. Most
  of the stars at large radii are carbon stars from the study of
  Kunkel \etal (1997); these stars are shown with larger symbols. A
  velocity gradient is visible by eye, and this is modeled in
  Section~\ref{s:los} to constrain rotation models for the LMC. The
  area shown in this figure is larger than that in
  Figures~\ref{f:obsvar}, \ref{f:obsrot}, and \ref{f:obscontour}.}
\label{f:losobs}
\end{center}
\end{figure*}

\subsection{Data}
\label{ss:losdata}

It is well-known that the kinematics of stars in the LMC depends on
the age of the population, as it does in the Milky Way. Young
populations have small velocity dispersions, and high rotation
velocities. By contrast, old populations have higher velocity
dispersions (e.g., van der Marel \etal 2009), and lower rotation
velocities (see Table~\ref{s:param}) due to asymmetric drift. For this
reason, we compiled two separate samples from the literature for the
present analysis: a ``young'' sample and an ``old'' sample. The young
sample is composed of RSGs, which is the youngest stellar population
for which detailed accurate kinematical data exist. The old sample is
composed of a mix of carbon stars, AGB stars, and RGB
stars\footnote{Many of these stars in the LMC are in fact
  ``intermediate-age'' stars, and are significantly younger than the
  age of the Universe. We use the term ``old'' for simplicity, and
  only in a relative sense compared to the younger RSGs.}.

For our young sample, we combined the RSG velocities of Prevot \etal
(1985), Massey \& Olsen (2003), and O11 (adopting the classification
from their figure~1). For the old sample, we combined the carbon star
velocities of Kunkel \etal (1997), Hardy et al.~(2001; as used also by
vdM02), and O11; the oxygen-rich and extreme AGB star velocities of
O11; and the RGB star velocities of Zhao \etal (2003; selected from
their figure~1 using the color criterion $B-R > 0.4$), Cole \etal
(2005), and Carrera \etal (2011).

When a star is found in more than one dataset, we retained only one of
the multiple velocity measurements. If a measurement existed from O11,
we retained that, because the O11 data set is the largest and most
homogeneous dataset available. Otherwise we retained the measurement
from the data set with the smallest random errors.

Stars with non-conforming velocities were rejected iteratively using
outlier rejection. For the young and old samples we rejected stars
with velocities that differ by more than $45 \kms$ and $90 \kms$ from
the best-fit rotation models, respectively. In each case this
corresponds to residuals $\gtrsim 4\sigma$, where $\sigma$ is the LOS
velocity dispersion of the sample. The outlier rejection removes both
foreground Milky Way stars, as well as stripped SMC stars that are
seen in the direction of the LMC (estimated by O11 as $\sim 6$\% of
their sample). 

All samples were brought to a common velocity scale by applying
additive velocity corrections to the data for each sample. These were
generally small\footnote{ Prevot \etal (1985): $+1.1 \kms$; Massey \&
  Olsen (2003): $+2.6 \kms$; Kunkel \etal (1997): $+2.7 \kms$; Hardy
  et al.~(2001): $-1.6 \kms$; Cole \etal (2005): $+3.0 \kms$; Carrera
  \etal (2011): $+2.5 \kms$.}, except for the Zhao \etal (2003)
sample\footnote{Field F056 Conf 01: $-16.6 \kms$; F056 Conf 02: $-6.2
  \kms$; F056 Conf 04: $-29.6 \kms$; F056 Conf 05: $-9.0 \kms$; F056
  Conf 21: $-16.8 \kms$; fields as defined in table~1 of Zhao \etal
  (2003).}. We adopted the absolute velocity scale of O11 as the
reference. Since they observed both young and old stars in the same
fields with the same setup, this ties together the velocity scales of
the young and old samples. To bring other samples to the O11 scale we
used stars in common between the samples, and we also compared the
residuals relative to a common velocity field fit.

Our final samples contain LOS velocities for 723 young stars and 6067
old stars in the LMC. Figure~\ref{f:losobs} shows a visual
representation of the discrete velocity field defined by the stars in
the combined sample. The coverage of the LMC is patchy and incomplete,
as defined by the observational setups used by the various studies.
The young star sample is confined almost entirely to distances
$\lesssim 4$ degrees from the LMC center. This is where the old star
sample has most of its measurements as well.  However, a sparse
sampling of old star velocities does continue all the way out to $\sim
14$ degrees from the LMC center. A velocity gradient is easily visible
in the figure by eye. What is observed is the sum of the internal
rotation of the LMC and an apparent solid-body rotation component due
to the LMC's transverse motion (vdM02). The latter component
contributes more as one moves further from the LMC center, which
causes an apparent twisting of the velocity field with radius.

\subsection{Fitting Methodology}
\label{ss:losmethod}

To interpret the LOS velocity data we use the same rotation field
model for a circular disk as in Section~\ref{ss:vmodel}. The model is
defined by the 7 parameters $(\alpha_0,\delta_0,D_0 \mu_{W0},D_0
\mu_{N0},v_{\rm LOS,0},i,\Theta)$ and the one-dimensional function
$V(R')$, which we parameterize with the two parameters $V_0$ and $R_0$
as in equation~(\ref{rotparam}). Note that the LOS velocity field
depends on the physical velocities $v_{W0} \equiv D_0 \mu_{W0}$,
$v_{N0} \equiv D_0 \mu_{N0}$, $v_{\rm LOS,0}$, and $V(R')$, unlike the
PM field, which depends on the angular velocities $\mu_{W0}$,
$\mu_{N0}$, $v_{\rm LOS,0}/D_0$, and $V(R')/D_0$.  As before, the
model can be written as a sum of two terms, $v_{\rm LOS,mod} = v_{\rm
  LOS,sys} + v_{\rm LOS,rot}$, representing the contributions from the
systemic motion of the LMC COM and from the internal rotation of the
LMC, respectively. The analytical expressions for the LOS velocity
field $v_{\rm LOS,mod}(\alpha,\delta)$ thus obtained were presented in
vdM02. As before, we refer the reader to that paper for the details of
the spherical trigonometry and linear algebra involved.

In Section~\ref{s:model} we have fit the PM data by themselves, and in
other studies such as vdM02 and O11, the LOS data have been fit by
themselves. These approaches require that some systemic velocity
components ($v_{\rm LOS,0}$ for the PM field analysis, and
$(\mu_{W0},\mu_{N0})$ for the LOS velocity field analysis) must be
fixed a priori to literature values. But clearly, the best way to use
the full information content of the data is to fit the PM and LOS data
{\it simultaneously}. This is therefore the approach we take here.

To fit the combined data, we define a $\chi^2$ quantity
\begin{equation}
  \chi^2 \equiv \chi^2_{\rm PM} + \chi^2_{\rm LOS}  .
\label{chitot}
\end{equation}
The quantity $\chi^2_{\rm PM}$ is as defined in
equation~(\ref{chisqpm}).  The observational PM errors are adjusted as
in Section~\ref{ss:compare} so that the best fit to the PM data by
themselves yields $\chi^2_{\rm PM} = N_{\rm DF}$. Similarly, we define
\begin{equation}
  \chi^2_{\rm LOS} \equiv 
  \sum_{i=1}^{N} \left [ (v_{{\rm LOS,obs},i} - v_{{\rm LOS,mod},i}) /
                         \sigma_{\rm LOS,obs} \right ]^2 ,
\label{chisqlos}
\end{equation}
which sums the squared residuals over all $N$ LOS velocities.  Here
$\sigma_{\rm LOS, obs}$ is a measure of the observed LOS velocity
dispersion of the sample, which we assume to be a constant for each
LOS velocity sample. We set $\sigma_{\rm LOS, obs}$ to be the RMS
scatter around the best-fit model that is obtained when the LOS data
are fit by themselves (this yields $\chi^2_{\rm LOS} = N$, 
analogous to the case for $\chi^2_{\rm PM}$). 


\begin{figure*}
\begin{center}
\epsfxsize=0.8\hsize
\centerline{\epsfbox{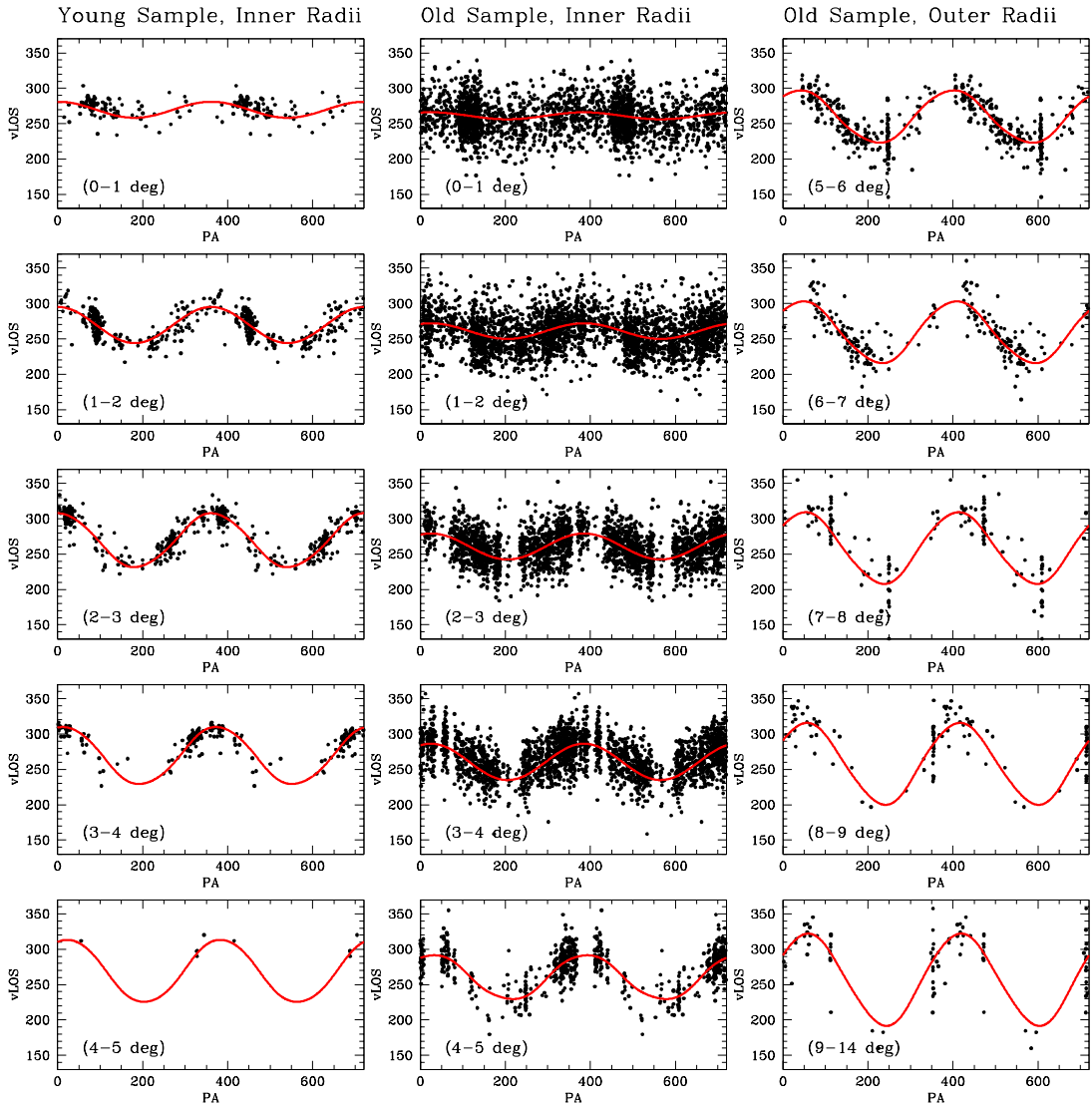}}
\caption{Data-model comparison for LOS velocities available from the
  literature. Each panel shows the heliocentric velocity of observed
  stars as function of the position angle $\Phi$ on the sky. The
  displayed range of the angle $\Phi$ is $0^{\circ}$--$720^{\circ}$,
  so each star is plotted twice. The left column is for the young star
  sample described in the text; the middle and right columns are for
  the old star sample. Each panel corresponds to a different range of
  angular distances $\rho$ from the LMC center, as indicated. The
  curves show the predictions of the best-fit models (calculated at
  the center of the radial range for the given panel), that also fit
  the new PM data.}
\label{f:obslos}
\end{center}
\end{figure*}

This approach yields that $\sigma_{\rm LOS,obs} = 11.6 \kms$ for the
young sample, and $\sigma_{\rm LOS,obs} = 22.8 \kms$ for the old
sample. This confirms, as expected, that the older stars have a larger
velocity dispersion. These results are broadly consistent with
previous work (e.g., vdM02; Olsen \& Massey 2007). Note that
$\sigma_{\rm LOS,obs}$ represents a quadrature sum of the intrinsic
velocity dispersion $\sigma_{\rm LOS}$ of the stars and the typical
observational measurement error $\Delta v_{\rm LOS}$. For all the data
used here, $\Delta v_{\rm LOS} \ll \sigma_{\rm LOS}$, so it is
justified to not include the individual measurement errors $\Delta
v_{{\rm LOS},i}$ explicitly in the definition of $\chi^2_{\rm LOS}$.

As before, we minimize $\chi^2$ as function of the model parameters
using a down-hill simplex routine (Press \etal 1992), with multiple
iterations and checks built in to ensure that a global minimum is
found. We calculate error bars on the best-fit model parameters using
Monte Carlo simulations. The pseudo PM data for this are generated as
in Section~\ref{ss:method}. The pseudo LOS velocity data are obtained
by drawing new velocities for the observed stars. For this we use the
predictions of the best-fit model, to which we add random Gaussian
deviates that have the same scatter around the fit as the observed
velocities.

In minimizing $\chi^2$, we treat all model parameters as free
parameters that are used to optimize the fit. However, we keep the
distance fixed at $m-M = 18.50$ (Freedman \etal 2001). The uncertainty
$\Delta (m-M) = 0.1$ is accounted for by including it in the
Monte-Carlo simulations that determine the uncertainties on the
best-fit parameters. As discussed later in Section~\ref{ss:dist}, the
combination of PM and LOS data does constrain the distance
independently. However, this does not (yet) yield higher accuracy than
conventional methods.

The stars for which we have measured PMs form essentially a magnitude
limited sample, composed of a mix of young and old stars. This mix is
expected to have a different rotation velocity than a sample composed
entirely of young or old stars. For this reason, we allow the rotation
amplitude $V_{0,{\rm PM}}$ in the PM field model to be different from
the rotation amplitude $V_{0,{\rm LOS}}$ in the velocity field
model. Both amplitudes are varied independently to determine the
best-fit model.  However, we do require the scale length $R_0$ of the
rotation curve and also the parameters that determine the orientation
and dynamical center of the disk to be the same for the PM and LOS
models.

With this methodology, we do two separate fits. The first fit is to
the combination of the PM data and the young LOS velocity sample, and
the second fit is to the combination of the PM data and the old LOS
velocity sample. This has the advantage (compared to a single fit to
all the data, with only a different rotation amplitude for each
sample) of providing two distinct answers. Comparison of the results
then provides insight into both the systematic accuracy of the
methodology, and potential differences in geometrical or kinematical
properties between different stellar populations.

\subsection{Data-Model Comparison}
\label{ss:loscompare}

Table~\ref{t:param} lists the parameters of the best-fit model and
their uncertainties. The quality of the model fits to the PM data is
similar to what was shown already in Figure~\ref{f:obsrot} for fits
that did not include any LOS velocity constraints. A data-model
comparison for the fits to the LOS velocity data is shown in
Figure~\ref{f:obslos}. The fits are adequate. It is clear that the
young stars rotate more rapidly than the old stars, and have a smaller
LOS velocity dispersion. The continued increase in the observed
rotation amplitude with radius is due to the solid-body rotation
component in the observed velocity field that is induced by the
transverse motion of the LMC.

The parameters for the best fit models to the combined PM and LOS
velocity samples can be compared to the results obtained when only the
PMs are fit (Table~\ref{t:param}), or the results that have been
obtained in the literature when only the LOS velocities were fit
(Table~\ref{t:paramlit}). This shows good agreement for some
quantities, and interesting differences for others. We proceed in
Section~\ref{s:param} by discussing the results and their comparisons
in detail, and what they tell us about the LMC.

\section{LMC Geometry, Kinematics, and Structure}
\label{s:param}


\begin{figure*}[t]
\begin{center}
\epsfxsize=0.8\hsize
\centerline{\epsfbox{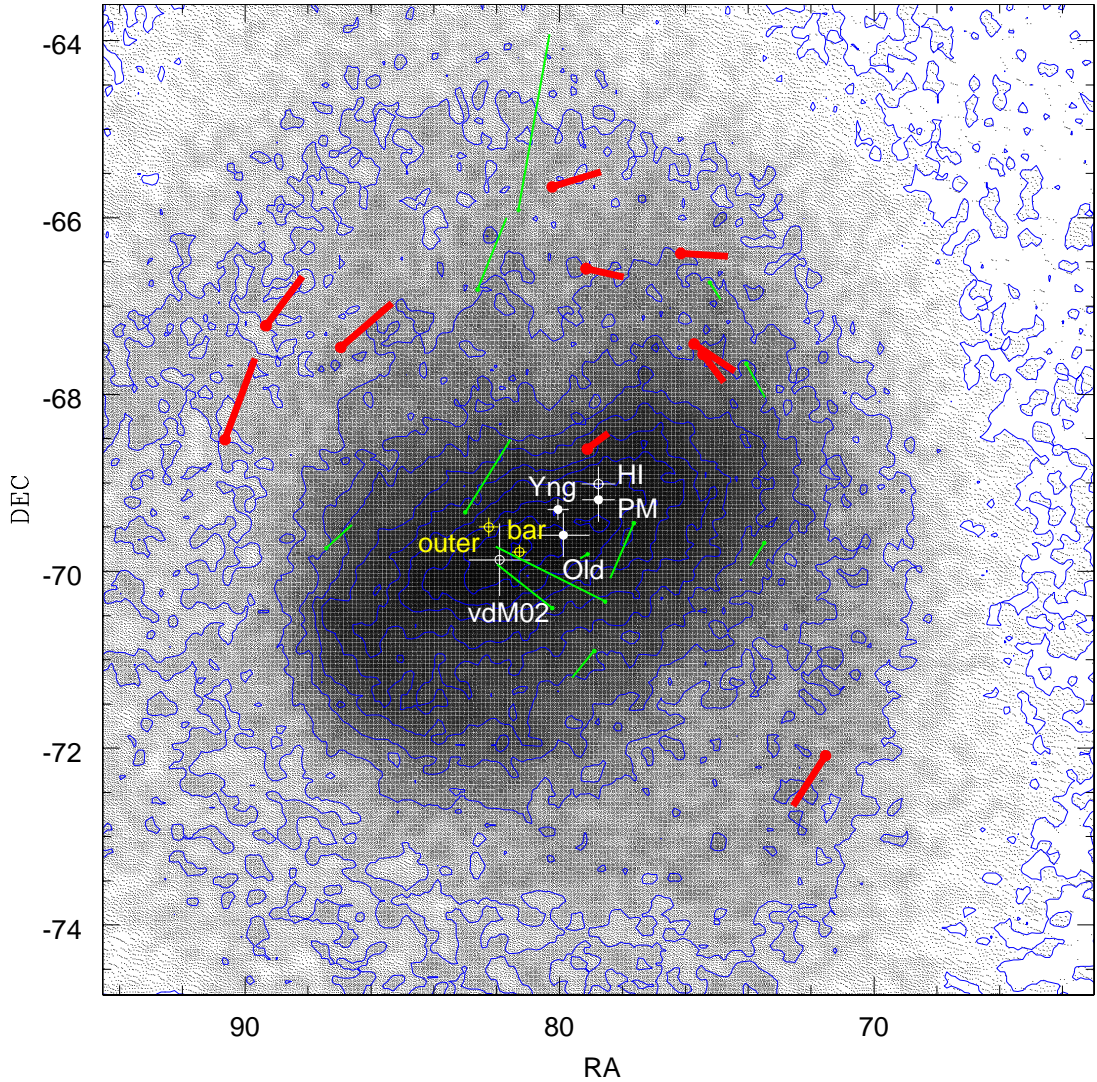}}
\caption{Determinations of dynamical and photometric centers of the
  LMC, overplotted on a grey-scale image with overlaid contours (blue)
  of the number density distribution of old stars in the LMC
  (extracted from the 2MASS survey; vdM01). Each center is discussed
  in the text, and is indicated as a circle with error bars. Solid
  circles are from the present paper (Table~\ref{t:param}), while open
  circles are from the literature. White circles are dynamical
  centers, while yellow circles are photometric centers. Labels are as
  follows. PM: stellar dynamical center inferred from the model fit to
  the new PM data; Old/Yng: stellar dynamical center inferred from the
  model fit to the combined sample of new PM data and old/young star
  LOS velocities; vdM02: stellar dynamical center previously inferred
  from the LOS velocity field of carbon stars; HI: gas dynamical
  center of the cold HI disk (Luks \& Rohlfs 1992; Kim \etal 1998);
  bar: densest point in the bar (de Vaucouleurs \& Freeman 1972;
  vdM01); outer: center of the outer isoplets, corrected for viewing
  perspective (vdM01). The rotation component ${\vec \mu}_{\rm
    obs,rot}$ of the observed LMC PM field is overplotted with similar
  conventions as in Figure~\ref{f:obsrot}. The three-epoch data (red)
  have significantly smaller uncertainties than the two-epoch data
  (green), but the actual uncertainties are shown only in
  Figure~\ref{f:obsrot}.}
\label{f:obscontour}
\end{center}
\end{figure*}

\subsection{Dynamical Center}
\label{ss:center}

The LMC is morphologically peculiar in its central regions, with a
pronounced asymmetric bar. Moreover, the light in optical images is
dominated by the patchy distribution of young stars and dust
extinction. As a result, the LMC has become known as a prototype of
``irregular'' galaxies (e.g., de Vaucouleurs \& Freeman
1972). However, the old stars that dominate the mass of the LMC show a
much more regular large-scale morphology. This is illustrated in
Figure~\ref{f:obscontour}, which shows the number density distribution
of red giant and AGB stars extracted from the 2MASS survey
(vdM01).\footnote{The figure shows a greyscale representation of the
  data in Figure~2c in vdM01, but in equatorial coordinates rather
  than a zenithal projection.} Despite this large-scale regularity,
there does not appear to be a single well-defined center. It has long
been known that different methods and tracers yield centers that are
not mutually consistent, as indicated in the figure.

The densest point in the LMC bar is located asymmetrically within the
bar, on the South-East side at $(\alpha_{\rm bar},\delta_{\rm bar}) =
(81.28^{\circ} \pm 0.24^{\circ}, -69.78^{\circ} \pm 0.08^{\circ})$
(vdM01; de Vaucouleurs \& Freeman 1972).\footnote{We
  adopt the center determined by vdM01, but base the
  error bar on the difference with respect to the center determined by
  de Vaucouleurs \& Freeman (1972). To facilitate comparison between
  different centers, we use decimal degree notation throughout for all
  positions, instead of hour, minute, second notation. The uncertainty
  in degrees of right ascension generally differs from the uncertainty
  in degrees of declination by approximately a factor $\cos(\delta)
  \approx 0.355$.} The center of the outer isoplets in
Figure~\ref{f:obscontour}, corrected for the effect of viewing
perspective, is at $(\alpha_{\rm outer}, \delta_{\rm outer}) =
(82.25^{\circ} \pm 0.31^{\circ}, -69.50^{\circ} \pm 0.11^{\circ})$
(vdM01). This is on the same side of the bar, but is
offset by $0.44^{\circ} \pm 0.14^{\circ}$. By contrast, the dynamical
center of the rotating HI disk of the LMC is on the opposite side of
the bar, at $(\alpha_{\rm HI}, \delta_{\rm HI}) = (78.77^{\circ} \pm
0.54^{\circ}, -69.01^{\circ} \pm 0.19^{\circ})$ (Kim \etal 1998; Luks
\& Rohlfs 1992).\footnote{We adopt the average of the centers
  determined by Kim \etal (1998) and Luks \& Rohlfs (1992), and
  estimate the error in the average based on the difference between
  these measurements.}  This is $1.18^{\circ} \pm 0.21^{\circ}$, i.e.,
more than 1 kpc, away from the densest point in the bar ($1 \kpc =
1.143^{\circ}$ at $D_0 = 50.1 \kpc$).

These offsets do not pose much of a conundrum. Numerical simulations
have established that an asymmetric density distribution and offset
bar in the LMC can be plausibly induced by tidal interactions with the
SMC (e.g., Bekki \etal 2009; Besla \etal 2012). What has been more
puzzling is the position of the stellar dynamical center at
$(\alpha_{\rm LOS}, \delta_{\rm LOS}) = (81.91^{\circ} \pm
0.98^{\circ}, -69.87^{\circ} \pm 0.41^{\circ})$, as determined by
vdM02 from the LOS velocity field of carbon stars. Olsen \& Massey
(2007) independently fit the same data, and obtained a position (and
other velocity field fit parameters) consistent with the vdM02 value.
The vdM02 stellar dynamical center was adopted by subsequent studies
of LOS velocities (e.g., O11) and PMs (K06, P08), without
independently fitting it. This position is consistent with the densest
point of the bar and with the center of the outer isophotes. But it is
$1.41^{\circ} \pm 0.43^{\circ}$ away from the HI dynamical
center. vdM02 argued that this may be due to the fact that HI in the
LMC is quite disturbed, and may be subject to non-equilibrium
gas-dynamical forces. However, more recent numerical simulations in
which the morphology of the LMC is highly disturbed due to
interactions with the SMC have shown that the dynamical centers of the
gas and stars often stay closely aligned (Besla \etal 2012).

The best-fit stellar dynamical center from our model fit to the PM
field is at $(\alpha_0, \delta_0) = (78.76^{\circ} \pm 0.52^{\circ},
-69.19^{\circ} \pm 0.25^{\circ})$. This {\it agrees} with the HI
dynamical center (see Figure~\ref{f:obscontour}). But it differs from
the stellar dynamical center inferred by vdM02 by $1.31^{\circ} \pm
0.44^{\circ}$, which is inconsistent at the 99\% confidence
level. This is surprising, because the PM field and LOS velocity field
are simply different projections of the three-dimensional velocity
field of the stellar population. So one would expect the inferred
dynamical centers to be the same.

When we fit the PM data and LOS velocities simultaneously
(Section~\ref{s:los}), we find centers that are somewhat intermediate
between between the PM-only dynamical center, and the vdM02 dynamical
center (see Figure~\ref{f:obscontour}). This is a natural outcome, as
these model fits try to compromise between datasets that apparently
prefer different centers. The old star sample that we use here is some
six times larger than the sample used by vdM02, and yields a center
that is consistent with the young star sample used here. Hence, the
fact that LOS velocities prefer a stellar dynamical center more
towards the South-East of the bar is a generic result, and does not
appear to be due to some peculiarity with the carbon star sample used
by vdM02. However, the dynamical centers that we infer from the
combined PM and LOS samples are much closer to the HI dynamical center
than the vdM02 dynamical center. Specifically, the offsets from the HI
center are $0.70^{\circ} \pm 0.33^{\circ}$ for the old $v_{\rm LOS}$
sample and $0.54^{\circ} \pm 0.22^{\circ}$ for the young $v_{\rm LOS}$
sample. Such offsets occur by chance only 9\% and 6\% of the time,
respectively. Hence, they most likely signify a systematic effect and
not just a chance occurrence.

In reality, it is likely that the HI and stellar dynamical centers are
coincident, since both the stars and the gas orbit in the same
gravitational potential. Some unknown systematic effect may therefore
be affecting the LOS velocity analyses. For example, there is good
reason to believe that the true dynamical structure of the LMC is more
complicated than the circular orbits in a thin plane used by our
models (e.g., warps and twists of the disk plane have been suggested
by vdMC01, Olsen \& Salyk 2002, and Nikolaev et al. 2004). The
uncertainties thus introduced may well affect different tracers
differently, leading to systematic offsets such as those reported
here. Visual inspection of the PM vector field in
Figure~\ref{f:obsrot} strongly supports that the center of rotation
must be close to the position identified by the PM-only model fit. For
example, the PM vectors in the central region do not have a definite
sense of rotation around the position identified by vdM02. Visual
inspection of the LOS velocity field in Figure~\ref{f:obslos} shows
the difficulty of determining an accurate center from such data.
Either way, the results in Table~\ref{t:param} and
Figure~\ref{f:obscontour} definitely indicate the LMC stellar
dynamical center is much closer to the HI dynamical center than was
previously believed.

\subsection{Disk Orientation}
\label{ss:orient}

Existing constraints on the orientation of the LMC disk come from two
techniques. The first technique is a geometric one, based on
variations in relative distance to tracers in different parts of the
LMC disk (vdMC01). The second is a kinematic method, based on fitting
circular orbit models to the velocity field of tracers, as we have done
here. The geometric technique yields both the inclination and
line-of-nodes position angle. When applied to LOS velocities, the
kinematic technique yields only the line-of-nodes position angle,
since the inclination is degenerate with the amplitude of the rotation
curve. But when applied to PMs, the kinematic technique yields both
viewing angles (see Section~\ref{ss:pmvslos}).

Existing constraints on the disk orientation obtained with these
techniques were reviewed in, e.g., van der Marel (2006) and van der
Marel \etal (2009). Some more recent results have appeared in e.g.
Koerwer (2009), O11, Haschke \etal (2012), Rubele \etal (2012), and
Subramanian \& Subramaniam (2013). All studies in the past decade or
so agree that the inclination is in the range $i \approx
25^{\circ}$--$40^{\circ}$, and that the line-of-nodes position angle
is in the range $\Theta \approx 120^{\circ}$--$155^{\circ}$. However,
the variations between the results from different studies are large,
and often exceed significantly the random errors in the best-fit
parameters. Some of this variation may be real, and due to spatial
variations in the viewing angles due to warps and twists of the disk
plane, combined with differences in spatial sampling between studies,
differences between different tracer populations, and contamination by
possible out of plane structures (e.g., O11).

Our best-fit model to the PM velocity field has $i = 39.6^{\circ} \pm
4.5^{\circ}$ and $\Theta = 147.4^{\circ} \pm 10.0^{\circ}$. The
implied viewing geometry of the disk is illustrated in
Figure~\ref{f:obsrot}. The inferred orientation angles are within the
range of expectation based on previous work, although they are at the
high end. However, they are perfectly plausible given what is known
about the LMC. This is an important validation of the accuracy of the
PM data and of our modeling techniques. It is the first time that PMs
have been used to derive the viewing geometry of any galaxy. However,
the random errors in our estimates are not sufficiently small to
resolve the questions left open by past work (apart from the fact that
variations in previously reported values appear to be dominated by
systematic variations, and not random errors).

When we fit not only the PM data, but also LOS velocities, the
best-fit viewing angles change (Table~\ref{t:param}), in some cases by
more than the random errors. However, all inferred values continue to
be within the range of what has been reported in the literature. The
best-fit inclination with PM data and the old star $v_{\rm LOS}$
sample is $i = 34.0^{\circ} \pm 7.0^{\circ}$, consistent e.g. with the
value $i = 34.7^{\circ} \pm 6.2^{\circ}$ inferred geometrically by
vdMC01 (and used subsequently in the kinematical studies of vdM02 and
O11). The best-fit line-of-nodes position angle with the PM data and
the old star $v_{\rm LOS}$ sample is $\Theta = 139.1^{\circ} \pm
4.1^{\circ}$. This is somewhat larger than the carbon star result
$\Theta = 129.9^{\circ} \pm 6.0^{\circ}$ obtained by vdM02, due
primarily to the different dynamical center inferred here.

The best-fit line-of-nodes position angle with the PM data and the
young star $v_{\rm LOS}$ sample is $\Theta = 154.5^{\circ} \pm
2.1^{\circ}$. This is larger than the result $\Theta = 142^{\circ} \pm
5^{\circ}$ obtained by O11 for the same $v_{\rm LOS}$ sample, due
primarily to the different dynamical center inferred here. The
best-fit inclination with the PM data and the young star $v_{\rm LOS}$
sample is $i = 26.2^{\circ} \pm 5.9^{\circ}$. This is somewhat smaller
than, but consistent with, the value obtained when the old star
$v_{\rm LOS}$ sample is used. However, the line-of-nodes position
angles for the fits with the young and old stars differ by $\Delta
\Theta = 15.4^{\circ} \pm 4.6^{\circ}$. This is an intriguing result,
since the data for these samples were analyzed in identical fashion,
and they do yield consistent dynamical centers. This suggests that
there may be real differences in the disk geometry or kinematics for
young and old stars, apart from their rotation amplitudes. Indeed, the
values inferred here kinematically using young stars are consistent
with the values inferred geometrically for (young) Cepheids, by
Nikolaev \etal (2004). They found that $i = 30.7^{\circ} \pm
1.1^{\circ}$ and $\Theta = 151.0^{\circ} \pm 2.4^{\circ}$. By
contrast, the values inferred here kinematically using old stars are
more consistent with some sets of orientation angles that have been
inferred geometrically for AGB and RGB stars (e.g., vdMC01; Olsen \&
Salyk 2002).

All results obtained here confirm once again that the position angle
of the line of nodes differs from the major axis of the projected LMC
body, which is at $189.3^{\circ} \pm 1.4^{\circ}$. This implies that
the LMC is not circular in the disk plane (vdM01).

\subsection{Systemic Transverse Motion}
\label{ss:LMCCOMPM}

In the best-fit model to the PM data, the final result for the LMC COM
PM is ${\vec \mu}_0 = (\mu_{W,0},\mu_{N,0}) = (-1.910 \pm 0.020, 0.229
\pm 0.047)$ mas/yr. Paper~I presented a detailed discussion of this
newly inferred value, including a comparison to previous HST and
ground-based measurements.

There are three components that contribute to the final PM error bars,
namely: (1) the random errors in the measurements of each field; (2)
the excess scatter between measurements from different fields that is
not accounted for by random errors, disk rotation, and viewing
perspective; and (3) uncertainties in the geometry and dynamics of the
best-fitting disk model. The contribution from the random errors can
be calculated simply by calculating the error in the weighted average
of all measurements. This yields $\Delta \mu_{W0,{\rm rand}} = \Delta
\mu_{N0,{\rm rand}} = 0.008$ mas/yr. This sets an absolute lower limit
to how well one could do in estimating the LMC COM PM from these data,
if there were no other sources of error. As discussed above, the
scatter between fields increases the error bars by a factor
$1.80$. Therefore, $\Delta \mu_{W0,{\rm rand+scat}} = \Delta
\mu_{N0,{\rm rand+scat}} = 0.014$ mas/yr. Since errors add in
quadrature, this implies that $\Delta \mu_{W0,{\rm scat}} = \Delta
\mu_{N0,{\rm scat}} = 0.012$ mas/yr. And finally the contribution from
uncertainties in geometry and dynamics of the best-fitting disk model
are $\Delta \mu_{W0,{\rm mod}} = 0.014$ mas/yr and $\Delta
\mu_{N0,{\rm mod}} = 0.045$ mas/yr. The final errors bars equal
$(\Delta \mu_{\rm rand}^2 + \Delta \mu_{\rm scat}^2 + \Delta \mu_{\rm
  mod}^2)^{1/2}$. So our knowledge of the geometry and kinematics of
the LMC disk is now the main limiting factor in our understanding of
the PM of the LMC COM.

The exact position of the LMC dynamical center is an important
uncertainty in models of the LMC disk. For this reason, we explored
explicitly how the fit to the PM velocity field depends on the assumed
center. For example, we ran models in which the center was kept fixed
to the position identified by vdM02 (even though this center is
strongly ruled out by our data). This changes only one of the COM PM
components significantly, namely $\mu_{N0}$, the LMC COM PM in the
North direction. Its value increases by $\sim 0.20$ mas/yr when the
vdM02 center is used instead of the best-fit PM center. When we use
instead the centers from our combined PM and LOS velocity fits, then
$\mu_{N0}$ increases by $0.06$--$0.10$ mas/yr, while again $\mu_{W0}$
stays the same to within the uncertainties (see Table~\ref{t:param}).
We have found more generally that if the center is moved roughly in
the direction of the position angle of the LMC bar (PA $\approx
115^{\circ}$; vdM01), then the implied $\mu_{N0}$ changes while the
implied $\mu_{W0}$ is unaffected. If instead the center is moved
roughly perpendicular to the bar, then $\mu_{W0}$ changes while the
implied $\mu_{N0}$ is unaffected. As discussed in Paper~I, $\mu_{W0}$
affects primarily the Galactocentric velocity of the LMC, while
$\mu_{N0}$ affects primarily the direction of the orbit as projected
on the sky. In practice, all of the centers that have been plausibly
identified for the LMC align roughly along the bar (see
Figure~\ref{f:obscontour}). Any remaining systematic uncertainties in
the LMC center position therefore affect primarily $\mu_{N0}$, and not
$\mu_{W0}$.

\subsection{Systemic Line-of-Sight Motion}
\label{ss:LMCCOMLOS}

In our fits to the PM field we kept the parameter $v_{\rm LOS,0}/D_0 =
1.104 \pm 0.053$ mas/yr fixed to the value implied by pre-existing
measurements. However, we did also run models in which it was treated
as a free parameter. This yielded $v_{\rm LOS,0}/D_0 = 1.675 \pm
0.687$ mas/yr. This is consistent with the existing knowledge, but not
competitive with it in terms of accuracy. Interestingly, the result
does show at statistical confidence that $v_{\rm LOS,0} > 0$. So the
observed PM field in Figure~\ref{f:obsvar} contains enough information
to demonstrate that the LMC is moving away from us. This is analogous
to the situation for the LOS velocity field, which contains enough
information to demonstrate that the LMC's transverse velocity is
predominantly directed Westward (figure~8 of vdM02).

In our fits of the combined PM and LOS velocity data, we did fit
independently for the systemic LOS velocity. When using the old star
$v_{\rm LOS}$ sample, this yields $v_{\rm LOS,0} = 261.1 \pm 2.2
\kms$. This is consistent with the results of vdM02 and Olsen \&
Massey (2007). However, when using the young star $v_{\rm LOS}$
sample, we obtain $v_{\rm LOS,0} = 269.6 \pm 1.9 \kms$. This differs
significantly both from the old star result, and from the result of
O11 for the same young star sample (Table~\ref{t:paramlit}). This is a
reflection of the different centers used in the various fits, and is
not due to an intrinsic offset in systemic velocity between young and
old stars. When we fit the young star data with a center that is fixed
to be identical to that for the old stars, we do find systemic
velocities $v_{\rm LOS,0}$ that are mutually consistent.
    
\subsection{Rotation Curve}
\label{ss:rotcurve}

\subsubsection{Rotation Curve from the Proper Motion Field}
\label{sss:pmrotcurve}

\begin{figure}
\begin{center}
\epsfxsize=\hsize
\centerline{\epsfbox{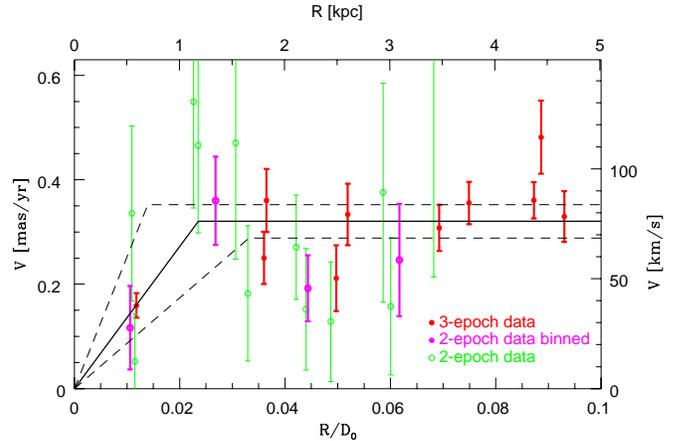}}
\caption{The LMC rotation curve inferred from the observed PM field as
  described in Section~\ref{sss:pmrotcurve}. $V$ is the rotation velocity
  in the disk at cylindrical radius $R$. The left and bottom axes are
  expressed in angular and dimensionless units, respectively, as
  directly constrained by the data. The right and top axes show the
  corresponding physical units, assuming an LMC distance $D_0 = 50.1
  \kpc$ ($m-M = 18.50$). Green and red data points show the results
  from individual HST fields with two and three epochs of data,
  respectively. Magenta data points show the result of binning the
  two-epoch data points into $R/D_0$ bins of size $0.018$. The red and
  magenta data points are listed in Table~\ref{t:rotcurve}. The black
  curve is the best-fit parameterization of the form given by
  equation~(\ref{rotparam}), with the surrounding black dashed curves
  indicating the $1\sigma$ uncertainty.}
\label{f:rotcurve}
\end{center}
\end{figure}

In the best-fit model to only the PM data, the rotation curve rises
linearly to $R_0/D_0 = 0.024 \pm 0.010$, and then stays flat at
$V_{0,{\rm PM}}/D_0 = 0.320 \pm 0.029$. At a distance modulus $m-M =
18.50 \pm 0.10$ (Freedman \etal 2001), this implies that $R_0 = 1.18
\pm 0.48 \kpc$ and $V_{0,{\rm PM}} = 76.1 \pm 7.6 \kms$. This rotation
curve fit is shown by the black lines in Figure~\ref{f:rotcurve}.

To further assess the PM rotation curve, we also obtained a
non-parametric estimate for it. For each HST field we already have
from Figure~\ref{f:obsrot} the observed rotation component ${\vec
  \mu}_{\rm obs,rot} \equiv {\vec \mu}_{\rm obs} - {\vec \mu}_0 -
{\vec \mu}_{\rm per}$, as well as the best-fit model component ${\vec
  \mu}_{\rm rot}$. The model also provides the in-plane rotation
velocity $V_{\rm mod}/D_0$ at the field location. We then estimate the
observed rotation velocity for each field as $V_{\rm obs}/D_0 =
[V_{\rm mod}(R)/D_0] ({\vec \mu}_{\rm obs,rot} \cdot {\vec \mu}_{\rm
  rot} / |{\vec \mu}_{\rm rot}|^2)$, where $\cdot$ designates the
vector inner product. This corresponds to modifying the model velocity
by the component of the residual PM vector that projects along the
local direction of rotation. Similarly, the uncertainty $\Delta V_{\rm
  obs}/D_0$ is estimated as the projection of the observational PM
error ellipse onto the rotation direction.

Figure~\ref{f:rotcurve} shows the rotation curve thus
obtained. Results are shown for the individual HST fields, color-coded
as in Figures~\ref{f:obsvar}, \ref{f:obsrot}, and~\ref{f:obscontour}
by whether two or three epochs of data are available. The three-epoch
measurements have good accuracy (median $\Delta V = 12 \kms$). By
contrast, the two-epoch measurements have much larger uncertainties
(median $\Delta V = 36 \kms$), as was the case in P08. So for the
two-epoch data we also plot the results obtained upon binning in $R'$
bins of size $0.018$. The rotation curve defined by combining the
three-epoch and binned two-epoch data is listed in
Table~\ref{t:rotcurve}. The unparameterized rotation curve is fit
reasonably well by the simple parameterization given by
equation~(\ref{rotparam}).


\begin{deluxetable*}{rrrrrrl}
\setlength{\tablewidth}{\hsize}
\tablecaption{LMC rotation curve from proper motions\label{t:rotcurve}}
\tablehead{
\colhead{$R/D_0$} & \colhead{$R$} & \colhead{$V/D_0$} & 
\colhead{$\Delta V/D_0$} & \colhead{$V$} & \colhead{$\Delta V$} &
\colhead{Field(s)}\\
\colhead{} & \colhead{(kpc)} & \colhead{(mas/yr)} & \colhead{(mas/yr)} &
\colhead{(km/s)} & \colhead{(km/s)} & \\
\colhead{(1)} & \colhead{(2)} & \colhead{(3)} & \colhead{(4)} & \colhead{(5)} &
\colhead{(6)} & \colhead{(7)} }
\startdata
0.0112 & 0.56 & 0.117 & 0.080 & 27.7 & 19.0 & L7,21 \\
0.0118 & 0.59 & 0.159 & 0.023 & 37.8 &  5.6 & L3 \\
0.0274 & 1.37 & 0.360 & 0.085 & 85.5 & 20.1 & L5,13,15,19 \\
0.0360 & 1.80 & 0.250 & 0.050 & 59.4 & 11.9 & L12 \\
0.0365 & 1.83 & 0.360 & 0.060 & 85.6 & 14.3 & L14 \\
0.0449 & 2.25 & 0.192 & 0.063 & 45.6 & 15.0 & L8,9,20 \\
0.0497 & 2.49 & 0.211 & 0.063 & 50.2 & 15.0 & L4 \\
0.0519 & 2.60 & 0.333 & 0.059 & 79.2 & 14.1 & L16 \\
0.0623 & 3.12 & 0.246 & 0.108 & 58.5 & 25.6 & L10,17,18 \\
0.0693 & 3.47 & 0.308 & 0.045 & 73.2 & 10.6 & L22 \\
0.0749 & 3.76 & 0.355 & 0.041 & 84.4 &  9.7 & L1 \\
0.0872 & 4.37 & 0.361 & 0.035 & 85.7 &  8.3 & L2 \\
0.0886 & 4.44 & 0.481 & 0.070 &114.4 & 16.6 & L6 \\
0.0930 & 4.66 & 0.330 & 0.049 & 78.3 & 11.6 & L11 \\
\enddata
\tablecomments{Column~(1) lists $R' \equiv R/D_0$, where $R$ is the
  radius in the disk. Column~(2) lists the corresponding $R$ in kpc,
  for an assumed LMC distance $D_0 = 50.1 \kpc$ ($m-M =
  18.50$). Column~(3) lists the rotation velocity $V/D_0$ in angular
  units, derived from the PM data as described in
  Section~\ref{ss:rotcurve}. Column~(4) lists the corresponding random
  uncertainty $\Delta V/D_0$. Columns~(5) and~(6) list the
  corresponding rotation velocity $V$ and its random uncertainty
  $\Delta V$ in km/s, for an assumed $D_0 = 50.1 \kpc$. Column~(7)
  lists the field identifiers from Paper~I. Three epoch measurements
  are listed singly, and two-epoch measurements are binned together in
  $R/D_0$ bins of size $0.018$. Errorbars include only the random noise
  in the measurements, and not the propagated errors from the
  uncertainties in other LMC model parameters. The rotation curve is shown in 
  Figure~\ref{f:rotcurve}.}
\end{deluxetable*}


P08 estimated the PM rotation curve from only the two-epoch PM
data. Their rotation velocity amplitude $V_{0,{\rm PM}} = 120 \pm 15
\kms$ was surprisingly high. This exceeds the value derived from the
radial velocities of HI and young stars by approximately 30--$40 \kms$
(O11). It would be hard to understand how any stars in the LMC could
be rotating significantly faster than the HI gas. When we use the
method discussed above on our own reanalysis of the two-epoch PM data,
the resulting unparameterized rotation curve is qualitatively similar
to that of P08, but the uncertainties are very large. With the
improved quality of our three-epoch data, the rotation curve is much
better determined. Moreover, the rotation amplitude comes down to
$V_{0,{\rm PM}} = 76.1 \pm 7.6 \kms$. This is more in line with
expectation, and removes a significant puzzle from the previous work.

The uncertainties in our unparameterized rotation curve in
Table~\ref{t:rotcurve} are underestimates, because they do not take
into account the propagated uncertainties in other model parameters
(such as the dynamical center, viewing angles, and COM motion). The
weighted average $V/D_0$ in Table~\ref{t:rotcurve} for $R > R_0$
equals $0.323 \pm 0.015$ mas/yr. By contrast, the best-fit from the
parameterized model in Table~\ref{t:param} is $0.320 \pm 0.029$
mas/yr. Since errors add in quadrature, the uncertainties in the other
model parameters contribute an uncertainty of $0.025$ mas/yr to the
rotation amplitude. This dominates the error budget, even though it is
small compared to the typical per-field error bars in
Table~\ref{t:rotcurve}. So the per-field PM uncertainties are not the
main limiting factor in our understanding of the rotation curve
amplitude.

When we fit not only the PM data, but also the LOS velocity data, then
the PM rotation amplitude $V_{0,{\rm PM}}$ changes by about the random
uncertainty $\Delta V_{0,{\rm PM}} = \pm 7.6 \kms$. When we fit the
old star $v_{\rm LOS}$ sample, $V_{0,{\rm PM}}$ goes up, and when we
fit the young star sample, $V_{0,{\rm PM}}$ goes down. This is because
the inclusion of the LOS velocities alters the best-fit line-of-nodes
position angle $\Theta$ to lower or higher values, respectively
(Table~\ref{t:param}). This affects $V_{0,{\rm PM}}$, because our
Monte-Carlo simulations show that $\Theta$ is anti-correlated with
$V_0$.

\subsubsection{Rotation Curve from the Line of Sight Velocity Field}
\label{sss:losrotcurve}

In our best-fit models to the $v_{\rm LOS}$ data (combined with the PM
data), the rotation amplitude $V_{0,{\rm LOS}}$ is not very accurately
determined. This is because only a fraction $\sin i$ of any velocity
is observed along the line of sight. The inclination is not accurately
known from our or any other data (see Section~\ref{ss:orient}), and
the deprojection therefore introduces significant uncertainty. By
contrast, $V_{0,{\rm LOS}}\>\sin i$ is determined much more
accurately. For our old star sample, we find that $V_{0,{\rm
    LOS}}\>\sin i = 30.9 \pm 2.6 \kms$. This is consistent with the
result from vdM02 (see Table~\ref{t:paramlit}). For our young star
sample, we find that $V_{0,{\rm LOS}}\>\sin i = 39.4 \pm 1.9 \kms$.
So the young stars have a higher rotation curve amplitude than the old
stars, consistent with previous findings. However, the value inferred
here is about 20\% less than the value implied by the rotation curve
fits of O11, for the same sample of stars (but not including PMs).
This is due primarily to the larger value of $\Theta$ inferred here.

\begin{figure}[t]
\begin{center}
\epsfxsize=\hsize
\centerline{\epsfbox{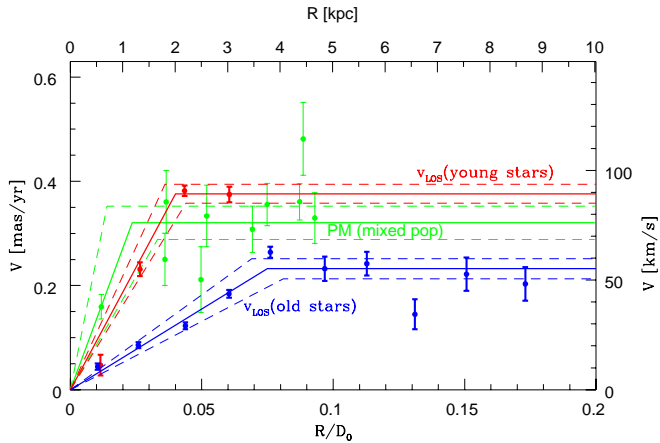}}
\caption{Comparison of LMC rotation curves inferred from different
  tracers as described in the text, with axes similar to
  Figure~\ref{f:rotcurve}. Red/Blue: young/old star $v_{\rm LOS}$
  sample from Table~\ref{t:rotcurveL}; Green: three-epoch HST PM data
  from Table~\ref{t:rotcurve}. Solid curves are the best-fit
  parameterizations of the form given by equation~(\ref{rotparam}),
  with parameters from Table~\ref{t:param}. The surrounding dashed
  curves indicate the $1\sigma$ uncertainty. Errorbars on the data
  points include only the shot/random noise from the measurements.
  The parameterized curves also include the propagated errors from the
  uncertainties in other LMC model parameters, except that the
  $v_{\rm LOS}$ fits shown do not include the inclination
  uncertainties.}
\label{f:rotcurveL}
\end{center}
\end{figure}

As for the PM case, we also determined unparameterized rotation curves
from the LOS velocity data, separately for the young and old star
samples. For this we kept all model parameters fixed, except the
rotation amplitude, to the values in Table~\ref{t:param}. We then
binned the stars by their radius $R' = R/D_0$ in the disk, and fit the
rotation amplitude separately for each radial bin. The rotation curves
thus obtained are listed in Table~\ref{t:rotcurveL}. The uncertainties
only take into account the shot noise from the finite number of
stars. This yields underestimates, because it does not take into
account the propagated uncertainties in other model parameters.  The
inferred rotation curves are shown in Figure~\ref{f:rotcurveL},
together with the parametrized fits from Table~\ref{t:param}. The
rotation curves are well fit by the simple parameterization given by
equation~(\ref{rotparam}). For the parameterized fits, the uncertainty
in the amplitude shown is $(\Delta V_{0,{\rm LOS}}\>\sin i) / \sin i$;
so this includes the propagated uncertainty from all model parameters
except the inclination. In general, for all rotation curve results
derived here from LOS velocities, the inclination is the dominant
uncertainty ($\Delta V / V = [\Delta i / 180^{\circ}] \pi / \tan i$).

The turnover radii in our rotation curve fits, $R_0/D_0 = 0.075 \pm
0.006$ for the old stars, and $R_0/D_0 = 0.040 \pm 0.004$ for the
young stars, are similar to what was found by vdM02 and O11,
respectively (Table~\ref{t:paramlit}). The value of $R_0$ for the
young stars is only about half that for the old stars. So the young
stars not only have a higher rotation curve amplitude, but the
rotation curve also rises faster. The value of the turnover radius
$R_0/D_0$ inferred from the fit to only the PM data, $R_0/D_0 = 0.024
\pm 0.010$, is even lower than the value for the young star $v_{\rm
  LOS}$ sample, but only at the $\sim 1.5\sigma$ level. We do not
attach much significance to this, given the sparse radial sampling of
our PM data, especially with high-quality WFC3 fields at small radii
(only one field at $R < 2.5$ kpc; see Figure~\ref{f:rotcurveL}). The
radial behavior and turnover of the rotation curve are therefore more
reliably constrained by the LOS data than by the PM data.

The values of $V_{0,{\rm LOS}}$ implied by our fits are $55.2 \pm
10.3$ for the old stars, and $89.3 \pm 18.8$ for the young stars,
respectively. These results are consistent with the results obtained
by vdM02 and O11 (Table~\ref{t:paramlit}). It should be noted that
while O11 reported $V_{0,{\rm LOS}} = 87 \pm 5 \kms$ for the same
sample of young stars, their listed uncertainty did not include the
uncertainty from propagation of uncertainties in the center,
inclination, COM PM, or distance. The inclination alone (from vdMC01,
as adopted by O11) adds a $14 \kms$ uncertainty. So while the random
uncertainties between our fit and that of O11 are in fact similar, our
result should be more accurate in a systematic sense. This is because
of our new determination of, e.g., the dynamical center and the COM
PM. The good agreement between the $V_{0,{\rm LOS}}$ values reported
here and in O11 is actually somewhat fortuitous. We find the LOS
component of the rotation to be $\sim 20$\% less than O11 did, but
they adopted a larger inclination.


\subsubsection{Comparison of Proper Motion and Line of Sight Rotation Curves}
\label{sss:rotcomp}

The rotation amplitude inferred from our PM data, $V_{0,{\rm PM}} =
76.1 \pm 7.6 \kms$, falls between the values inferred from the LOS
velocities of old stars, $V_{0,{\rm LOS}} = 55.2 \pm 10.3 \kms$, and
young stars, $V_{0,{\rm LOS}} = 89.3 \pm 18.8 \kms$, respectively (see
also Figure~\ref{f:rotcurveL}). This is because our stellar PM sample
is essentially a magnitude limited sample, composed of a mix of young
and old stars.

To assess quantitatively whether the rotation amplitudes derived from
the PM data and LOS velocities are consistent, let us assume that a
fraction $f$ of the stars that contribute to our PM measurements are
young, and a fraction $(1-f)$ are old. The designations ``young'' and
``old'' in this context refer to the fact that the stars are assumed
to have the same kinematics as the stars in our young and old $v_{\rm
  LOS}$ samples. This implies an expected PM-inferred rotation
amplitude $V_{0,{\rm PM}} = 55.2 + f(34.1) \pm \sqrt([(1-f)10.3]^2 +
[f18.8]^2) \kms$. Equating this with the observed $V_{0,{\rm PM}}$
implies that $f = 0.61 \pm 0.42$.


\begin{deluxetable}{cccccc}
\setlength{\tablewidth}{\hsize}
\tablecaption{LMC rotation curve from LOS velocities\label{t:rotcurveL}}
\tablehead{
\colhead{$R/D_0$} & \colhead{$R$} & 
\colhead{$V$} & \colhead{$\Delta V$} &
\colhead{$V$} & \colhead{$\Delta V$} \\
 & & \colhead{[young]} & \colhead{[young]} & 
     \colhead{[old]} & \colhead{[old]} \\ 
\colhead{} & \colhead{(kpc)} & \colhead{(km/s)} & \colhead{(km/s)} & 
\colhead{(km/s)} & \colhead{(km/s)} \\
\colhead{(1)} & \colhead{(2)} & \colhead{(3)} & \colhead{(4)} & \colhead{(5)} &
\colhead{(6)} }
\startdata
0.011 & 0.5 & 11.2     & 4.7      & 10.6 & 1.4 \\
0.026 & 1.3 & 54.9     & 3.1      & 20.3 & 1.4 \\
0.044 & 2.2 & 90.7     & 2.4      & 29.2 & 1.6 \\
0.060 & 3.0 & 89.0     & 3.6      & 43.7 & 1.7 \\
0.076 & 3.8 & $\ldots$ & $\ldots$ & 62.7 & 2.5 \\
0.097 & 4.9 & $\ldots$ & $\ldots$ & 55.2 & 5.5 \\
0.113 & 5.7 & $\ldots$ & $\ldots$ & 57.4 & 5.4 \\
0.131 & 6.6 & $\ldots$ & $\ldots$ & 34.4 & 6.8 \\
0.151 & 7.6 & $\ldots$ & $\ldots$ & 52.7 & 7.6 \\
0.173 & 8.7 & $\ldots$ & $\ldots$ & 48.3 & 7.6 \\
\enddata
\tablecomments{Column~(1) lists $R' \equiv R/D_0$, where $R$ is the
  radius in the disk. Column~(2) lists the corresponding $R$ in kpc,
  for an assumed LMC distance $D_0 = 50.1 \kpc$ ($m-M =
  18.50$). Columns~(3) and~(4) list the rotation velocity $V$ in km/s
  with its uncertainty, determined as in
  Section~\ref{sss:losrotcurve}, for the young $v_{\rm LOS}$
  sample. Columns~(5) and~(6) list the same quantities for the old
  $v_{\rm LOS}$ sample. Error bars include only the shot noise from the
  measurements, and not the propagated errors from the uncertainties
  in other LMC model parameters. The rotation curves are shown in
  Figure~\ref{f:rotcurveL}.}
\end{deluxetable}

Figure~6 of K06 shows a color-magnitude diagram (CMD) of the LMC stars
that contribute to our PM measurements. At the magnitudes of interest,
there are two main features in this diagram. There is a blue plume,
consisting of main sequence stars and evolved massive stars at the
blue edge of their blue loops. And there is a red plume, consisting
mostly of RGB stars and some AGB stars. Bright RSGs, such as those in
the $v_{\rm LOS}$ samples, are too rare to contribute significantly to
our small HST fields. To count the relative numbers of blue and red
stars, we adopt a separation at $V-I = 0.65$. We then find that the
fraction of blue stars (as ratio of the total stars that contribute to
our PM measurements) increases from $\sim 50$\% at the brightest
magnitudes to $\sim 70$\% at the faintest magnitudes. If we assume
that the blue stars have kinematics typical of young stars, and the
red stars have kinematics typical of old stars, then this implies $f
\approx 0.6 \pm 0.1$. This CMD-based value is in excellent agreement
with the value inferred above from the observed kinematics. So to
within the uncertainties, the observed rotation of the LMC in the PM
direction is consistent with the observed rotation in the LOS
direction.

The LMC rotation amplitude $V_{0,{\rm PM}}$ inferred from the PM field
is relatively insensitive to the inclination (see
Section~\ref{ss:pmvslos}). By contrast, the LOS velocity data
accurately constrain $V_{0,{\rm LOS}}\>\sin i$. Since the fraction $f$
{\it must} be between 0 and 1, comparison of these quantities can set
limits on the inclination. The inclination must be such that
$[V_{0,{\rm LOS}}\>\sin i]_{\rm old} / \sin i \leq V_{0,{\rm PM}} \leq
[V_{0,{\rm LOS}}\>\sin i]_{\rm young} / \sin i$. With the inferred
values from Table~\ref{t:param} this implies that at $1\sigma$
confidence $18.5^{\circ} \leq i \leq 39.3^{\circ}$. As discussed in
Section~\ref{ss:orient}, this encompasses most of the results reported
in the literature.  Alternatively, we could be less conservative and
assume that we know from the CMD analysis that $f = 0.6 \pm 0.1$. In
that case we obtain the more stringent range $24.3^{\circ} \leq i \leq
32.4^{\circ}$. But this assumes that we know the difference in
kinematics between different stars in our HST CMDs, which has not
actually been measured.

\subsection{Kinematical Distance Estimates}
\label{ss:dist}

So far, we have assumed that the distance $D_0$ to the LMC center of
mass is known. However, a comparison of the PM and LOS velocity fields
does in fact constrain the distance independently, since PMs are
measured in mas/yr, and LOS velocities are measured in km/s. As we will
discuss, this comparison provides several independent distance
constraints.

The first distance constraint is obtained by requiring that the
rotation amplitude measured from PMs matches that obtained from LOS
velocities. This is called the ``rotational parallax'' method. Based
on the discussion in the previous section, this implies that
\begin{equation}
   D_0 = ( f [V_{0,{\rm LOS}}]_{\rm young} +
           (1-f) [V_{0,{\rm LOS}}]_{\rm old} ) / [V_{0,{\rm PM}} / D_0] .
\end{equation} 
To use this equation, we must assume the relative fractions of young
and old stars that contribute to the PM measurements. Using the
analysis in Section~\ref{sss:rotcomp}, we set $f=0.6 \pm 0.1$. With
the inferred values from Table~\ref{t:param} this implies that $D_0 =
18.48 \pm 0.40$. This is consistent with existing knowledge (e.g.,
Freedman \etal 2001). However, the uncertainty is very large, due
primarily to the uncertainties in the LMC inclination. To obtain a
distance estimate with a random error $\Delta (m-M) \leq 0.1$, the
inclination would have to be known to better than $1.5^{\circ}$, not
even accounting for other uncertainties. Based on the discussion in
Section~\ref{ss:orient}, it is clear that this is not currently the
case, despite many papers devoted to the subject. 
Moreover, one would need to know the fraction $f$
more accurately than is possible with only CMD information. So for the
LMC, the method of rotational parallax is not likely to soon yield a
competitive distance estimate.

An alternative method to constrain the LMC distance from comparison of
the PM and LOS velocity fields uses the observed LOS velocities
perpendicular to the line of nodes. Rotation is perpendicular to the
line of sight there, so that the observed velocities are due entirely
to the solid-body rotation induced by the LMC's transverse
motion. Hence, the velocities obey $v_{\rm LOS} = \pm D_0 \mu_{\perp}
\sin \rho$, where $\rho$ is the distance from the COM, and
$\mu_{\perp}$ is the component of the COM PM that lies along the line
of nodes (vdM02). Since $\mu_{\perp}$ is constrained by the PM data in
mas/yr, the distance $D_0$ can be determined from the LOS data in
km/s. For accurate results, this method benefits from having data that
extends to large distances $\rho$, and from having a sample with many
velocity measurements. We therefore apply it to the old star $v_{\rm
  LOS}$ sample (see Figure~\ref{f:obslos}). To use the full
information content of the data, and to adequately propagate all
uncertainties, one must fit the combined PM and old star $v_{\rm LOS}$
sample with $m-M$ as a free parameter. When we do this while keeping
$\Theta = 139.1^{\circ} \pm 4.1^{\circ}$ fixed to the previously
obtained value from Table~\ref{t:param}, we obtain $m-M = 18.53 \pm
0.20$ (similar to an earlier estimate in van der Marel \etal (2009),
which was based on the vdM02 carbon star LOS velocity data and the K06
COM PM estimate). This has a smaller random error than the result from
the rotational parallax method, but is still not competitive with
existing knowledge. Moreover, it may be a biased estimate. When
fitting $m-M$, one should really fit $\Theta$ simultaneously, because
$\Theta$ and $m-M$ are generally anti-correlated in our model
fits. However, we found that the multi-dimensional solution space
becomes more degenerate when both $\Theta$ and $m-M$ are left to
vary. Specifically, the best fit $m-M$ can vary by $\pm 0.2$,
depending on how we choose to weight the PM data relative to the LOS
velocity data in the $\chi^2$ definition (eq.~[\ref{chitot}]). So this
method does not currently yield a competitive distance either.

A final method for estimating the the LMC distance from comparison of
the PM and LOS velocity fields uses the observed systemic LOS
velocity. As stated in Section~\ref{ss:LMCCOMLOS}, our PM field fit
constrains $v_{\rm LOS,0}/D_0 = 1.675 \pm 0.687$ mas/yr. Using the
known systemic LOS velocity $v_{\rm LOS,0} = 261.1 \pm 2.2 \kms$ for
the old star $v_{\rm LOS}$ sample, this yields an estimate for the
distance: $m-M = 17.58 \pm 0.89$. Again, this is consistent with
existing knowledge, but not competitive in terms of accuracy.

\subsection{Disk Precession and Nutation}
\label{ss:precess}

The preceding analysis in this paper has assumed that the viewing
angles of the LMC disk are constant with time. vdM02 showed that there
are additional contributions to PM and LOS velocity fields when the
viewing angles vary with time, i.e., $di/dt \not= 0$ or $d\Theta/dt
\not=0$. This corresponds to a precession or nutation of the spin axis
of the LMC disk. To induce such motion requires external tidal
torques. While it is not impossible that such motion may exist, there
is no theoretical requirement that it should.

The main impact of a value $di/dt \not= 0$ is to induce a solid-body
rotation component in the LOS velocity field, with its steepest
gradient perpendicular to the line of nodes. To assess the existence
of such a component, we repeated our fits to the combined PM data and
old star $v_{\rm LOS}$ sample, but now with $di/dt$ free to vary.
This yields $di / dt = -0.08 \pm 0.17$ mas/yr. This result is
consistent with zero. So with the presently available data, there is
no need to invoke a non-zero value of $di / dt$.

Constraints on $di/dt$ from the young star $v_{\rm LOS}$ sample are
weaker, because those data don't extended as far from the COM, and
don't have as many velocity measurements. O11 inferred $di / dt =
-0.66 \pm 0.29$ mas/yr for that same sample ($-184^{\circ} \pm
81^{\circ}$ Gyr$^{-1}$). However, their uncertainty is an
underestimate, because it does not propagate the known uncertainties
in the center, inclination, COM PM, or distance. Based on our analysis
of the young stars, we have found no compelling reason to assume they
require $di / dt \not =0$. It would in fact be difficult to understand
how the spin axis of the young star disk could be moving relative to
the old star disk.

A value $d\Theta/dt \not=0$ does not affect the LOS velocity
field. However, it does cause circular motion in the observed PM
field. This is almost entirely degenerate with the actual rotation of
the LMC disk, as measured by the rotation amplitude $V_{0,{\rm PM}}$
(compare Figure~\ref{f:obsrot}). We have shown in
Section~\ref{sss:rotcomp} that the amplitude inferred assuming
$d\Theta/dt =0$ agrees with the rotation amplitudes inferred from LOS
velocities. Therefore, the data do not require a value $d\Theta/dt
\not=0$. If there is a deviation from zero, it would have to be small
enough to not perturb the agreement discussed in
Section~\ref{sss:rotcomp}.

\subsection{Mass}
\label{ss:mass}

To estimate the dynamical mass of the LMC, it is necessary to know the
kinematics of tracers at large radii. The outermost tracers for which
kinematics are available are the old stars in our $V_{\rm LOS}$ sample
(see Figure~\ref{f:obslos}), most of which are carbon stars from
Kunkel \etal (1997; see Figure~\ref{f:losobs}).
Figure~\ref{f:rotcurveL} shows that the rotation curve of these stars
stays more-or-less flat out the last data point, at radius $R = 8.7
\kpc$ in the disk (see also figure~6 of vdM02). Since this is true for
the old stars, it must be true for the young stars as well. After all,
both orbit in the same gravitational potential.

Based on this reasoning, we infer that the young stars have a rotation
amplitude $V_{0,{\rm LOS}} = 89.3 \pm 18.8 \kms$ at $R = 8.7
\kpc$. Olsen \& Massey (2007) inferred a velocity dispersion for these
stars of $\sigma_{\rm LOS} = 9 \kms$. The formalism of vdM02 then
implies an upward asymmetric drift correction of only $\Delta V = 2.4
\kms$, much smaller than the random errors. This is as expected, given
that O11 found that the young stars and HI gas have essentially the
same rotation curve. So we obtain that $V_{\rm circ} = 91.7 \pm 18.8
\kms$ at $R = 8.7$ kpc, with the error dominated by inclination
uncertainties.

The total mass of the LMC inside the last measured data point is
$M_{\rm LMC} (R) = R V_{\rm circ}^2 / G$, where the gravitational
constant $G = 4.3007 \times 10^{-6} \kpc (\kms)^2 \Msun^{-1}$. This
yields $M_{\rm LMC} (8.7 \kpc) = (1.7 \pm 0.7) \times 10^{10} \Msun$.
This is consistent with the total mass of the LMC that has been used
in several past $N$-body simulation studies of the LMC (e.g., Gardiner
\& Noguchi 1996). However, this is likely an underestimate of the
total LMC mass. Since the rotation curve is flat, the mass likely
continues to rise almost linearly beyond $8.7 \kpc$. Many virial
masses are possible, as long as the concentration of the dark halo is
varied to reproduce the dynamical mass. In Paper~I we considered
models with virial masses up to $25 \times 10^{10} \Msun$.

\subsection{Tidal Radius}
\label{ss:tidal}

The tidal radius of the LMC can be estimated using the formalism of
vdM02. We assume that the LMC rotation curve is flat at $V_{\rm circ}
= 91.7 \pm 18.8 \kms$, and the Milky Way rotation curve at the
distance of the LMC is flat at $V_{\rm MW} = 206 \pm 23 \kms$ (based
on the enclosed mass out to the LMC distance given by Kochanek
1996). This implies a tidal radius of $22.3 \pm 5.2 \kpc$ (i.e., a
radius on the sky of $24.0^{\circ} \pm 5.6^{\circ}$).  If instead
there is no LMC mass outside of $8.7 \kpc$, then the tidal radius is
smaller by a factor $0.73$. Either way, the LMC tidal radius is beyond
$17^{\circ}$. Indeed, photometric studies of the LMC have traced the
LMC disk almost this far out (Saha \etal 2010).

\subsection{Tully-Fisher Relation}
\label{ss:cosmo}

\begin{figure}
\begin{center}
\epsfxsize=\hsize
\centerline{\epsfbox{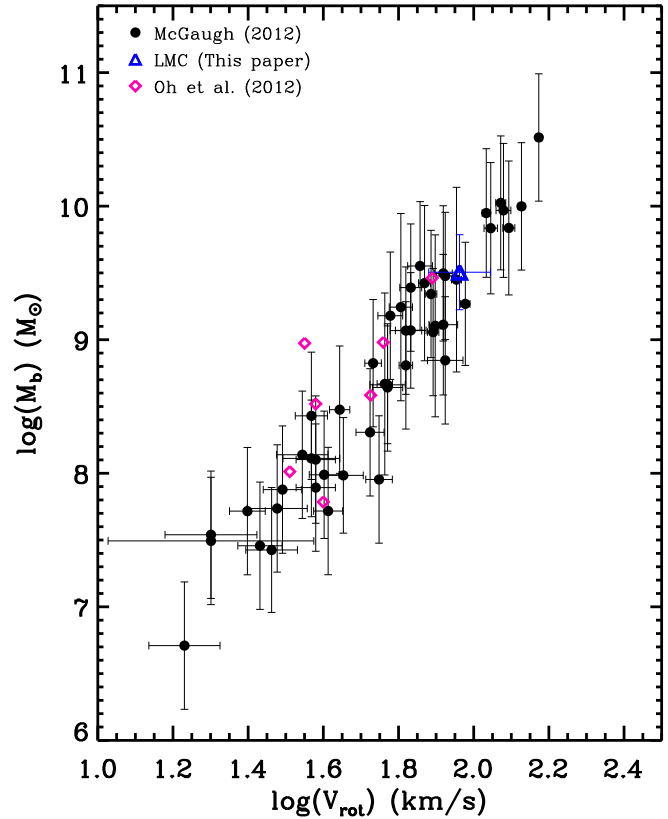}}
\caption{Baryonic galaxy mass versus rotation velocity in a log-log
  plot. The LMC (blue triangle) follows the baryonic Tully-Fisher
  relation defined by low-mass gas-dominated galaxies (black circles;
  McGaugh 2012) and dwarf galaxies from the THINGS survey (pink
  diamonds; Oh \etal 2012a,b).}
\label{f:BTF}
\end{center}
\end{figure}

To determine whether the rotation curve of the LMC is typical, it is
useful to assess how its circular velocity compares to that of other
similar galaxies. The classical Tully-Fisher relation in spirals has
been shown to extend into the low-mass regime when the total baryonic
content of the galaxies is used (gas in addition to stars; McGaugh
\etal 2005; Stark \etal 2009). In Figure~\ref{f:BTF} (blue triangle)
we place the LMC on the baryonic Tully-Fisher (BTF), using $V_{\rm
  circ}$ from Section~\ref{ss:mass} and $M_{\rm b} = 3.2 \times 10^9
\Msun$ from vdM02.


McGaugh (2012) recently calibrated the BTF relation using a sample of
gas-dominated low-mass systems, arguing that the errors introduced by
modeling the stellar component is minimized in these gas-rich
systems. These galaxies follow a very tight relation. As with
high-mass galaxies, the scatter is below what is expected from initial
conditions in cosmological models, implying the need for some feedback
process that is correlated with the galaxy potential (Eisenstein \&
Loeb 1996; McGaugh 2012). Similarly, Oh \etal (2011a) placed a sample
of dwarf galaxies from the HI Nearby Galaxy Survey (THINGS; Walter et
al.~2008), a high velocity-resolution H{\small I} survey of dwarf
galaxies, on the BTF relation. In Figure~\ref{f:BTF}, black points
show gas-rich low-mass galaxies compiled in McGaugh (2012) and pink
diamonds show the THINGS dwarfs from Oh \etal (2011a,b). The LMC falls
on the BTF relation followed by these galaxies. So even though the LMC
is a member of an interacting pair, it is not atypical in terms of its
BTF position.

As discussed in Section 4.8, the error in our $V_{\rm circ}$ is
dominated by inclination uncertainties. So if one assumes {\it a
  priori} that the LMC must fall exactly on the BTF, then this would
in principle provide an alternative method to constrain the LMC
inclination (see also Oh \etal 2011a). Also, central density profile
slopes are a powerful probe of the properties of dark matter (Navarro,
Frenk, \& White \etal 1997; Dalcanton \& Hogan 2001), and these can be
constrained from observed rotation curves of dwarf galaxies (e.g., de
Blok 2010, and references therein; Oh \etal 2011a,b). However, for the
LMC only sparse discrete PM and LOS datasets are available, and these
are not ideally suited for constraining the central rotation curve
slope. Also, the possibility of non-circular orbits in the poorly
understood LMC bar region would complicate any interpretation. So we
have decided not to pursue a detailed rotation curve decomposition
here.

\section{Conclusions}
\label{s:conc}

We have presented a detailed study of the large-scale rotation of the
LMC based on observed stellar velocities in all three Cartesian
directions. This is the first time that such a study has been possible
for any galaxy, made possible by the exquisite capabilities of HST for
measuring PMs in the nearby Universe. While the LOS velocity field of
the LMC has been studied previously using many tracers, our analysis
of the PM rotation field is new. This is important, because the PM
rotation field is defined by two components of motion, and it
therefore has a higher information content than the LOS velocity
field. As a result, quantities that are degenerate in analyses of the
LOS velocity field (such as the rotation curve, the inclination, and
one component of the transverse motion of the COM) are uniquely
determined by analysis of the PM rotation field. We interpret the data
with simple models of circular rotation in a flat disk, which fit the
data reasonably well. By and large, we find that the LMC rotation
properties as revealed by PM and LOS data are mutually consistent, as
they should be. However, by analyzing accurate PM data and combining
it with existing LOS data we do obtain several new insights into the
geometry, kinematics, and structure of the LMC.

Previous studies of the LMC have found that the photometric center is
offset significantly from the dynamical center defined by the rotating
HI gas disk. This is not difficult to explain, since the LMC has a
lopsided off-center bar that could be a transient feature induced by
the LMC's interaction with the SMC. What is more puzzling has been the
finding that the LOS velocity field of stellar tracers is best fit by
a dynamical center that is {\it also} offset from the HI dynamical
center.  Our new analysis of the PM rotation field does not confirm
this. We find that the stellar dynamical center revealed by PMs {\it
  agrees} with the HI dynamical center. However, we also find that
previous analyses of the LOS velocity field were not in error. Our new
analysis of the now very large sample of available LOS velocity data
continues to indicate a dynamical center offset, albeit by a smaller
amount than reported previously. This cannot be real, since the PM and
LOS analyses observe the same stellar populations. This likely reveals
limitations of the simple rotation model used. In reality, the stars
and gas in the LMC probably do have the same dynamical center, because
they orbit in the same gravitational potential.

The best-fit values for the viewing angles that define the orientation
of the LMC disk, as inferred from the PM rotation field, are within
the range of values implied by previous studies. However, several
puzzles remain. First, the position angle of the line of nodes
$\Theta$ is not the same for the young and old stellar populations of
the LMC. When LOS velocities of the young population are fit jointly
with the PM data, we obtain $\Theta = 154.5^{\circ} \pm 2.1^{\circ}$.
By contrast, when LOS velocities of the old population are fit jointly
with the PM data, we obtain $\Theta = 139.1^{\circ} \pm 4.1^{\circ}$.
When the PM data are fit by themselves, the intermediate result
$\Theta = 147.4^{\circ} \pm 10.0^{\circ}$ is obtained. The second
puzzle is that all these kinematically determined values are larger
than several results obtained from geometrical methods (e.g., vdMC01,
Rubele \etal 2012). Similarly for the inclination, the results
obtained here and those discussed in the literature span a much larger
range than the random errors in the individual measurements. These
results can be explained if the structure of the LMC is more
complicated than a single flat disk in circular rotation. Indeed, the
data provide indications for variations with both stellar population
and radius in the disk. However, by contrast to previous authors we
have found no evidence for precession or nutation of the LMC
disk. Given the latest insights into the position and motion of the
LMC COM, we find that acceptable fits to all the kinematical data can
be obtained with $d i / dt = 0$ and $d \Theta / dt = 0$.

The LMC rotation curve as implied by our PM measurements has an
amplitude $V_{0,{\rm PM}} = 76.1 \pm 7.6 \kms$. This applies to a
magnitude-limited sample, composed of a mix of stellar populations.
This value of $V_{0,{\rm PM}}$ falls between the rotation amplitudes
implied by the LOS velocities of old and young stars, $V_{0,{\rm LOS}}
= 55.2 \pm 10.3$ and $89.3 \pm 18.8$, respectively. These results are
quantitatively consistent with the natural hypothesis that the blue
stars in our HST CMDs have predominantly young-star kinematics and the
red stars have predominantly old-star kinematics. These results
resolve a puzzle posed by analysis of the two-epoch PM rotation curve
amplitude. P08 previously reported $V_{0,{\rm PM}} = 120 \pm 15 \kms$,
which was difficult to understand as it exceeded the rotation
amplitude of both the young stars and the HI gas.

After correction for asymmetric drift, we infer a circular velocity
for the LMC of $V_{\rm circ} = 91.7 \pm 18.8 \kms$. The large
uncertainty is due mostly to uncertainties in the inclination of the
LMC. This $V_{\rm circ}$ places the LMC on the same baryonic
Tully-Fisher relation defined by samples of other low-mass gas-rich
galaxies. Also, it implies an enclosed mass $M_{\rm LMC} = (1.7 \pm
0.7) \times 10^{10} \Msun$, out to the radius $8.7 \kpc$ to which it
has been verified that the rotation curve remains approximately
flat. The virial mass of the LMC should be larger than this, with the
exact value depending on how far the LMC's dark halo extends, and on
whether it has been tidally truncated. If the LMC circular velocity
curve remains flat outside of the region probed observationally, then
the tidal radius is $22.3 \pm 5.2 \kpc$ (i.e., a radius on the sky of
$24.0^{\circ} \pm 5.6^{\circ}$).

We have discussed three independent methods for combining PM and LOS
velocity information to obtain a kinematical estimate for the LMC
distance. While each of these methods yields results that are
consistent with existing knowledge, none of them is currently
competitive in terms of accuracy. This is due in large part to the
fact that the exact viewing angles of the LMC continue to be poorly
understood.  These distance determination methods might become
competitive in the future, if better PM data become available and our
understanding of the structure and orientation of the LMC improve
further.

\acknowledgements

Support for this work was provided by NASA through grants for program
GO-11730 from the Space Telescope Science Institute (STScI), which is
operated by the Association of Universities for Research in Astronomy
(AURA), Inc., under NASA contract NAS5-26555. The authors are grateful
to Gurtina Besla, Jay Anderson, and Charles Alcock for contributing to
the original observing proposals and to other papers in this series.
Knut Olsen kindly made available in electronic format the data
presented in O11. NK would like to thank Marla Geha's research group
for valuable discussions.

{\it Facilities:} \facility{HST (ACS/HRC; WFC3/UVIS)}.


{}

\end{document}